%% file: main.tex
\documentclass[twocolumn,tighten]{aastex631}
\input{affil}
\usepackage{CJK}

\begin{document}

\title{The Early Light Curve of SN 2023bee: Constraining Type Ia Supernova Progenitors the Apian Way}

\correspondingauthor{Griffin Hosseinzadeh}
\email{griffin0@arizona.edu}

\author[0000-0002-0832-2974]{Griffin Hosseinzadeh}
\UA
\author[0000-0003-4102-380X]{David J.\ Sand}
\UA
\author[0000-0002-4781-7291]{Sumit K.\ Sarbadhicary}
\OSU
\author[0000-0003-4501-8100]{Stuart D.\ Ryder}
\Macquarie\AAARC
\author[0000-0001-8738-6011]{Saurabh W.\ Jha}
\Rutgers
\author[0000-0002-7937-6371]{Yize Dong \begin{CJK*}{UTF8}{gbsn}(董一泽)\end{CJK*}\!\!}
\UCD
\author[0000-0002-4924-444X]{K.\ Azalee Bostroem}
\Catalyst\UA
\author[0000-0003-0123-0062]{Jennifer E.\ Andrews}
\GeminiNorth
\author[0000-0003-2744-4755]{Emily Hoang}
\UCD
\author[0000-0003-0549-3281]{Daryl Janzen}
\USask
\author[0000-0001-5754-4007]{Jacob E.\ Jencson}
\UA
\author[0000-0001-9589-3793]{Michael Lundquist}
\Keck
\author[0000-0002-7015-3446]{Nicolas E.\ Meza Retamal}
\UCD
\author[0000-0002-0744-0047]{Jeniveve Pearson}
\UA
\author[0000-0002-4022-1874]{Manisha Shrestha}
\UA
\author[0000-0001-8818-0795]{Stefano Valenti}
\UCD
\author[0000-0003-2732-4956]{Samuel Wyatt}
\UW
\author[0000-0003-4914-5625]{Joseph Farah}
\LCO\UCSB
\author[0000-0003-4253-656X]{D.\ Andrew Howell}
\LCO\UCSB
\author[0000-0001-5807-7893]{Curtis McCully}
\LCO\UCSB
\author[0000-0001-9570-0584]{Megan Newsome}
\LCO\UCSB
\author[0000-0003-0209-9246]{Estefania Padilla Gonzalez}
\LCO\UCSB
\author[0000-0002-7472-1279]{Craig Pellegrino}
\LCO\UCSB
\author[0000-0003-0794-5982]{Giacomo Terreran}
\LCO
\author{Muzoun Alzaabi}
\UA
\author{Elizabeth M.\ Green}
\UA
\author{Jessica L.\ Gurney}
\UA
\author[0000-0002-0370-157X]{Peter A.\ Milne}
\UA
\author{Kaycee I.\ Ridenhour}
\UA
\author[0000-0001-5510-2424]{Nathan Smith}
\UA
\author{Paulina Soto Robles}
\UA
\author[0000-0003-3108-1328]{Lindsey A.\ Kwok}
\Rutgers
\author[0009-0002-5096-1689]{Michaela Schwab}
\Rutgers
\author[0000-0002-1650-1518]{Mariusz Gromadzki}
\Warsaw
\author[0000-0002-7004-9956]{David A.\ H.\ Buckley}
\SAAO\UFS\UCT
\author{Koichi Itagaki \begin{CJK*}{UTF8}{min}(板垣公一)\end{CJK*}\!\!}
\Itagaki
\author[0000-0002-1125-9187]{Daichi Hiramatsu}
\CfA\IAIFI
\author[/0000-0002-8400-3705]{Laura Chomiuk}
\MSU
\author[0000-0002-3664-8082]{Peter Lundqvist}
\OKC
\author[0000-0002-6703-805X]{Joshua Haislip}
\UNC
\author[0000-0003-3642-5484]{Vladimir Kouprianov}
\UNC
\author[0000-0002-5060-3673]{Daniel E.\ Reichart}
\UNC

\submitjournal{\textit{The Astrophysical Journal Letters}}
\received{2023 May 4}
\revised{2023 June 30}
\accepted{2023 July 12}

\begin{abstract}
We present very early photometric and spectroscopic observations of the Type~Ia supernova (SN~Ia) 2023bee, starting about 8~hr after the explosion, which reveal a strong excess in the optical and nearest UV ($U$ and \textit{UVW1}) bands during the first several days of explosion. This data set allows us to probe the nature of the binary companion of the exploding white dwarf and the conditions leading to its ignition. We find a good match to the Kasen model in which a main-sequence companion star stings the ejecta with a shock as they buzz past. Models of double detonations, shells of radioactive nickel near the surface, interaction with circumstellar material, and pulsational delayed detonations do not provide good matches to our light curves. We also observe signatures of unburned material, in the form of carbon absorption, in our earliest spectra. Our radio nondetections place a limit on the mass-loss rate from the putative companion that rules out a red giant but allows a main-sequence star. We discuss our results in the context of other similar SNe~Ia in the literature.
\end{abstract}

\keywords{Binary stars (154), Supernovae (1668), Type Ia supernovae (1728), White dwarf stars (1799)}

\section{Introduction} \label{sec:intro}
Despite their utility as cosmological distance indicators, the physics behind Type~Ia supernovae (SNe~Ia), the explosions of carbon--oxygen white dwarfs, is still poorly understood (see \citealt{hoeflich_explosion_2017} for a review). In the single-degenerate scenario, the white dwarf accretes material from a main-sequence or giant star \citep{whelan_binaries_1973}, whereas the double-degenerate scenario involves a binary system of two white dwarfs \citep{iben_supernovae_1984,webbink_double_1984}. In addition, the explosion may be a delayed detonation \citep{khokhlov_delayed_1991} or a double detonation \citep{nomoto_accreting_1982,livne_successive_1990,woosley_sub-chandrasekhar_2011}, and the latter may or may not be dynamically driven \citep{shen_three_2018}. These models have various, sometimes overlapping, observational signatures (see \citealt{jha_observational_2019} for a recent review).

With the advent of high-cadence time-domain surveys, the early light curves (starting $\lesssim$1 day after explosion) of SNe~Ia have been of particular interest as probes of their progenitor systems and explosion mechanisms. \cite{kasen_seeing_2010} predicted that the collision between SN ejecta and a nondegenerate binary companion could lead to a blue excess in the first few days after explosion for ${\sim}10\%$ of viewing angles. Similar bumps can be produced by interaction with circumstellar material \citep[CSM;][]{piro_exploring_2016}, unusual distributions of radioactive $^{56}$Ni \citep{noebauer_early_2017}, and the double-detonation mechanism itself \citep{polin_observational_2019}. However, when examined in detail, these models make different predictions for the strength, duration, and color of the early excess.

The first well-sampled early SN~Ia light curve was of SN~2011fe, showing a smooth parabolic rise in $g$ band over the first several days after explosion \citep{nugent_supernova_2011}. However, over the past decade, several SNe~Ia have shown unusual behavior in various filter bands immediately following explosion, ranging from a very fast rise to an initial decrease in luminosity. Normal to overluminous (99aa-like) SNe~Ia with early blue or UV excesses include SN~2012cg \citep{marion_sn_2016}, SN~2017cbv \citep{hosseinzadeh_early_2017}, iPTF16abc \citep{miller_early_2018}, SN~2021aefx \citep{ashall_speed_2022,hosseinzadeh_constraining_2022,ni_origin_2023}, and SN~2021hpr \citep{lim_early_2023}. Underluminous SNe~Ia with strong UV excesses include iPTF14atg \citep{cao_strong_2015} and SN~2019yvq \citep{miller_spectacular_2020,siebert_strong_2020,burke_bright_2021,tucker_sn_2021}. MUSSES16D4 \citep{jiang_hybrid_2017} and SN~2018aoz \citep{ni_infant-phase_2022,ni_origin_2023a}, on the other hand, show early red excesses. Normal SN~Ia~2018oh \citep{dimitriadis_k2_2019,li_photometric_2019,shappee_seeing_2019} and overluminous (03fg-like) SN~Ia~2021zny \citep{dimitriadis_sn_2023} also show early excesses in their high-cadence unfiltered light curves from Kepler~2 and the Transiting Exoplanet Survey Satellite (TESS), respectively, although the color of this excess is not known. An excess can be ruled out in three other SNe~Ia with equivalent data from Kepler \citep{olling_no_2015}. It is not clear which, if any, of these groups are physically related. Sample analyses by \cite{burke_companion_2022,burke_early_2022} and \cite{deckers_constraining_2022} show excesses in ${\sim}10\%$ of SNe~Ia with early photometry.

Here we present observations of SN~2023bee, another overluminous (99aa-like) SN~Ia discovered within $\sim$1 day of explosion and observed at very high cadence by the Distance Less Than 40~Mpc (DLT40) Survey \citep{tartaglia_early_2018}. We describe these observations in \S\ref{sec:obs}. The early light curve shows an excess in the $U$ band during the first $\sim$3 days after explosion, which we fit with various models in \S\ref{sec:phot}. We analyze the early spectra of SN~2023bee in \S\ref{sec:spec} and use our radio observations to place constraints on CSM in \S\ref{sec:radioanalysis}. In \S\ref{sec:discuss}, we discuss the implications of our analysis on the progenitor systems of SNe~Ia in general and SN~2023bee in particular.

\section{Observations and Data Reduction} \label{sec:obs}

\subsection{Discovery and Classification} \label{sec:disc}
The DLT40 Survey discovered SN~2023bee on 2023-02-01.75 UT at an unfiltered brightness of $17.26 \pm 0.04$~mag and did not detect it the previous night (2023-01-31.12) to an unfiltered limit of $>$19.68~mag \citep{andrews_dlt40_2023}. The SN occurred at J2000 coordinates $\alpha=08\textsuperscript{h}56\textsuperscript{m}11\fs63$ and $\delta=-03\degr19'32\farcs1$, $135''$ north--northeast of the center of NGC~2708, which has a redshift of $z=0.006698 \pm 0.000017$ \citep{falco_updated_1999}. \cite{zhai_lions_2023} initially classified it as a Type~Ic SN, but it was reclassified as an SN~Ia by \cite{hosseinzadeh_global_2023} based on a spectrum taken the following night.

We also recovered an earlier marginal detection of SN~2023bee in a pair of unfiltered images taken on 2023-01-31.65 with Koichi Itagaki's Bitran BN-83M CCD imager mounted on a 0.5~m telescope in Okayama Prefecture, Japan. We aligned and stacked these images using Astrometry.net \citep{lang_astrometry_2010} and \texttt{reproject} \citep{robitaille_image_2023}, and measured aperture photometry using Photutils \citep{bradley_astropy_2022}. We calibrated this measurement to $r$-band magnitudes from the Pan-STARRS1 $3\pi$ Survey \citep{chambers_pan-starrs1_2016}. The resulting magnitude is $18.7 \pm 0.5$, which is significant at the ${\sim}2\sigma$ level. Given the positional coincidence, we assume the detection is real and include this point in our analysis.

TESS was actively observing the location of SN~2023bee at the time of its explosion. Unfortunately, severe contamination from scattered light within $\sim$1 day of explosion make those images unusable for photometry, but the remainder of the TESS light curve is analyzed by \cite{wang_flight_2023}.

\subsection{Photometry}

\begin{figure*}
    \centering
    \includegraphics[width=\textwidth]{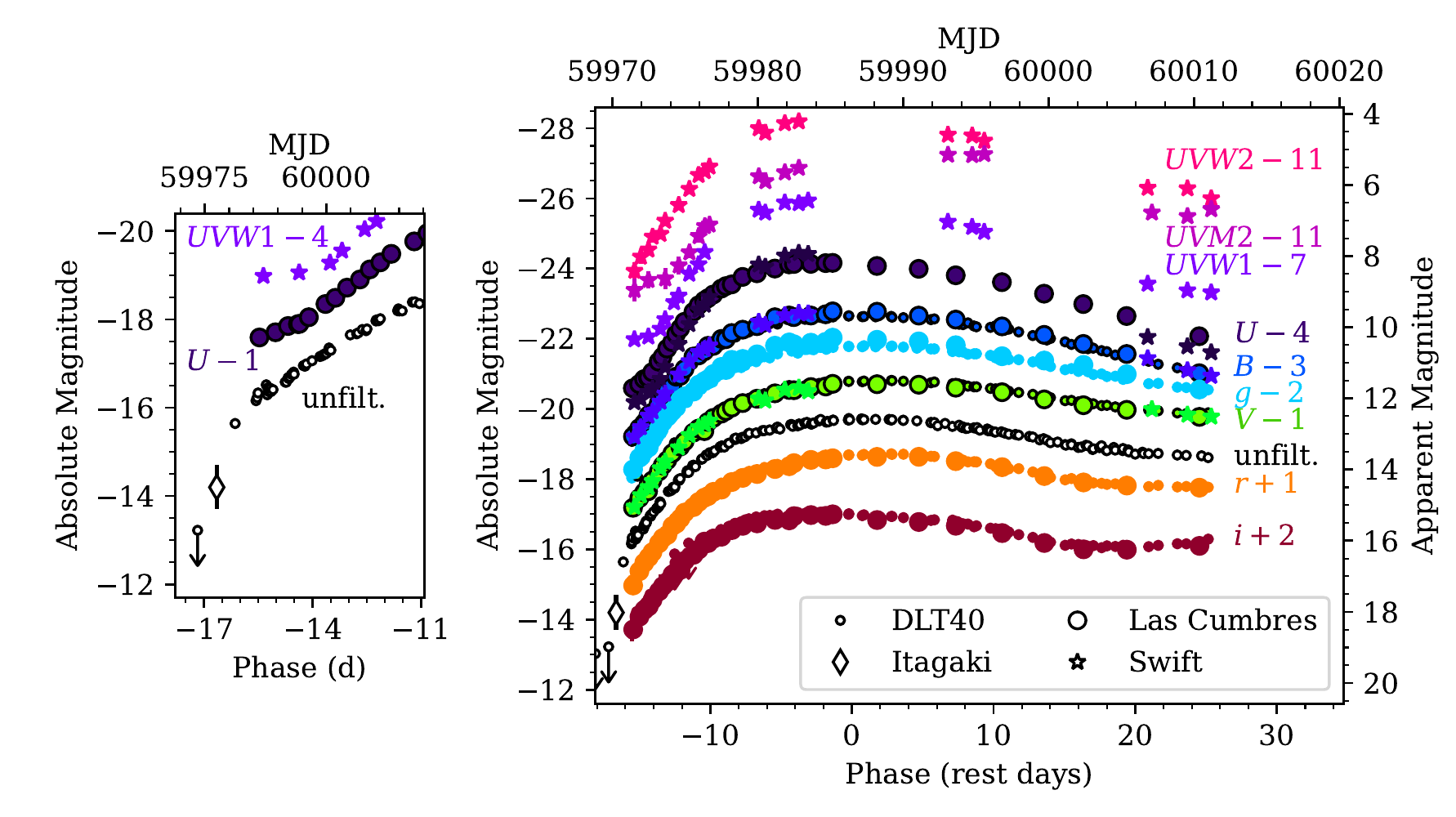}
    \caption{The multiband light curve of SN~2023bee in absolute and extinction-corrected apparent magnitudes. Phases are in rest-frame days with respect to the $B$-band peak. Downward-pointing arrows indicate $3\sigma$ nondetections. No offset is applied to the unfiltered photometry. The left panel shows the very early DLT40 discovery 1 day after a deep nondetection. The diamond shows a $2\sigma$ detection by Koichi Itagaki. During this period, the $U$-band (purple) and \textit{UVW1}-band (violet) light curves show a distinct change in slope, indicating the end of an excess with respect to a smooth rise. \\ (The data used to create this figure are available.)}
    \label{fig:phot}
\end{figure*}

After discovery, we initiated a high-cadence optical--UV follow-up campaign using the Sinistro cameras on Las Cumbres Observatory's network of 1~m telescopes \citep{brown_cumbres_2013} as part of the Global Supernova Project, the Panchromatic Robotic Optical Monitoring and Polarimetry Telescopes \citep{reichart_prompt:_2005} as part of the DLT40 Survey, and the Ultraviolet/Optical Telescope on the Neil Gehrels Swift Observatory \citep{roming_swift_2005}. We measured point-spread function photometry on the Las Cumbres images and aperture photometry on the other images using dedicated pipelines based on PyRAF \citep{sciencesoftwarebranchatstsci_pyraf:_2012}, HEAsoft \citep{nasaheasarc_heasoft:_2014}, and Photutils \citep{bradley_astropy_2022}. Las Cumbres \textit{UBV} photometry is calibrated to \cite{rubin_second_1974} and \cite{landolt_ubvri_1983,landolt_ubvri_1992} standard fields observed on the same nights at the same sites. Las Cumbres \textit{gri} photometry is calibrated to the Pan-STARRS1 $3\pi$ Catalog \citep{chambers_pan-starrs1_2016}. DLT40 photometry is calibrated to the AAVSO Photometric All-Sky Survey \citep{henden_aavso_2009}, with unfiltered photometry calibrated to the $r$ band. Swift photometry uses the updated zero-points of \cite{breeveld_further_2010} with the time-dependent sensitivity corrections from 2020. \textit{UBV} and Swift magnitudes are in the Vega system, and \textit{gri} magnitudes are in the AB system. Figure~\ref{fig:phot} shows the full multiband light curve, which is available in machine-readable form.

\subsection{Spectroscopy}

\begin{figure*}
    \centering
    \includegraphics[width=0.95\textwidth]{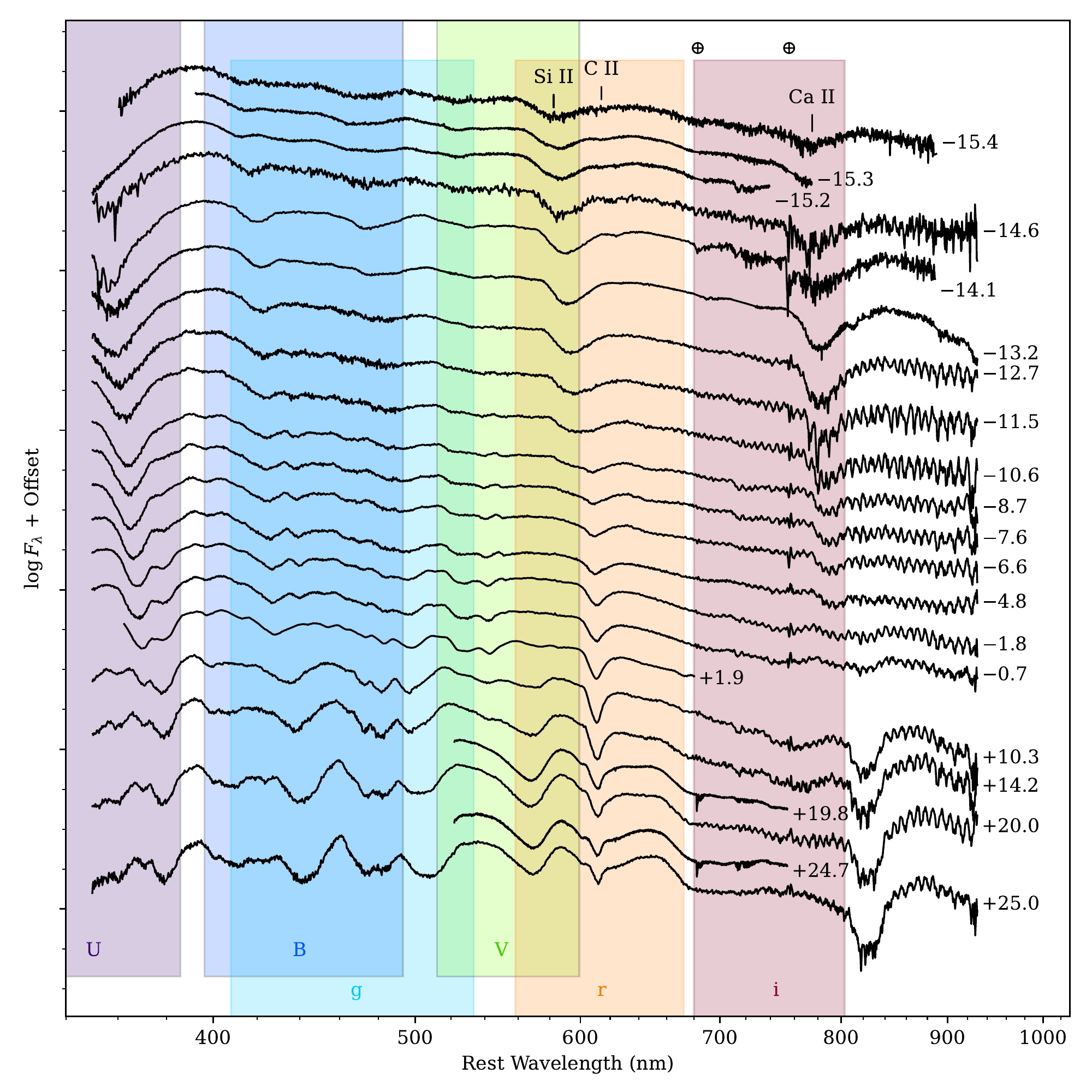}
    \caption{Spectral series of SN~2023bee, with phases with respect to the $B$-band peak shown at the right. The colored bands represent the optical bandpasses discussed in the text. Major absorption features due to \ion{Si}{2}, \ion{C}{2}, \ion{Ca}{2}, and telluric lines ($\oplus$) are marked at the top. The observation of carbon early in the evolution implies the presence of unburned material in the ejecta. Our very high-cadence early spectroscopy shows a rapid decrease in photospheric velocity and an initial increase in equivalent width (see Table~\ref{tab:spec}). The latter may indicate an early hot blackbody component that fades after a few days. \\ (The data used to create this figure are available.)}
    \label{fig:spec}
\end{figure*}

\begin{deluxetable*}{cccccccR}
\tablecaption{Log of Spectroscopic Observations, Velocities, and Equivalent Widths} \label{tab:spec}
\tablehead{\colhead{} & \colhead{} & \colhead{} & \multicolumn2c{\ion{Si}{2} 636 nm} & \multicolumn2c{\ion{C}{2} 658 nm} & \colhead{Phase} \\[-12pt]
\colhead{MJD} & \colhead{Telescope} & \colhead{Instrument} & \multicolumn2c{---------------------------------------} & \multicolumn2c{---------------------------------------} & \colhead{} \\[-10pt]
\colhead{} & \colhead{} & \colhead{} & \colhead{Vel. (km s$^{-1}$)} & \colhead{EW (nm)} & \colhead{Vel. (km s$^{-1}$)} & \colhead{EW (nm)} & \colhead{(d)}}
\startdata
59976.763 & LJT & YFOSC    & 25,220 $\pm$ 160 & \phd9.8 $\pm$ 0.9 & 22,960 $\pm$ 520 & 3.6 $\pm$ 1.0 & -15.4 \\
59976.885 & SALT & RSS     & 24,520 $\pm$ \phd80 & 10.7 $\pm$ 1.5 & 21,510 $\pm$ 250 & 3.4 $\pm$ 1.6 & -15.3 \\
59976.985 & SALT & RSS     & 23,980 $\pm$ \phd80 & 13.0 $\pm$ 1.4 & 20,790 $\pm$ 260 & 4.1 $\pm$ 1.5 & -15.2 \\
59977.564 & FTS & FLOYDS   & 23,050 $\pm$ \phd90 & 12.3 $\pm$ 1.0 & 21,440 $\pm$ 670 & 2.7 $\pm$ 1.1 & -14.6 \\
59978.115 & SOAR & Goodman & 21,740 $\pm$ \phd90 & 13.4 $\pm$ 1.1 & 20,540 $\pm$ 600 & 2.9 $\pm$ 1.2 & -14.1 \\
59979.002 & SALT & RSS     & 21,050 $\pm$ \phd60 & 14.1 $\pm$ 1.2 & \nodata & \nodata & -13.2 \\
59979.513 & FTS & FLOYDS   & 20,200 $\pm$ \phd90 & 13.1 $\pm$ 0.8 & \nodata & \nodata & -12.7 \\
59980.705 & FTS & FLOYDS   & 18,880 $\pm$ 110 & 10.8 $\pm$ 0.8 & \nodata & \nodata & -11.5 \\
59981.641 & FTS & FLOYDS   & 17,930 $\pm$ 170 & 11.6 $\pm$ 0.7 & \nodata & \nodata & -10.6 \\
59983.512 & FTS & FLOYDS   & 15,470 $\pm$ 190 & \phd8.6 $\pm$ 0.8 & \nodata & \nodata & -8.7 \\
59984.630 & FTS & FLOYDS   & 13,710 $\pm$ 170 & \phd7.7 $\pm$ 0.7 & \nodata & \nodata & -7.6 \\
59985.648 & FTS & FLOYDS   & 13,150 $\pm$ 110 & \phd6.2 $\pm$ 0.8 & \nodata & \nodata & -6.6 \\
59987.492 & FTS & FLOYDS   & 12,490 $\pm$ \phd80 & \phd5.7 $\pm$ 0.9 & \nodata & \nodata & -4.8 \\
59990.502 & FTS & FLOYDS   & 12,070 $\pm$ \phd60 & \phd6.0 $\pm$ 0.8 & \nodata & \nodata & -1.8 \\
59991.604 & FTS & FLOYDS   & 12,150 $\pm$ \phd50 & \phd6.1 $\pm$ 0.8 & \nodata & \nodata & -0.7 \\
59994.208 & Bok & B\&C     & 12,210 $\pm$ \phd50 & \phd6.8 $\pm$ 0.6 & \nodata & \nodata & +1.9 \\
60002.620 & FTS & FLOYDS   & 12,220 $\pm$ \phd40 & \phd7.5 $\pm$ 0.6 & \nodata & \nodata & +10.3 \\
60006.598 & FTS & FLOYDS   & 11,970 $\pm$ \phd40 & \phd7.4 $\pm$ 0.6 & \nodata & \nodata & +14.2 \\
60012.205 & MMT & Binospec & 11,710 $\pm$ \phd30 & \phd5.3 $\pm$ 1.2 & \nodata & \nodata & +19.8 \\
60012.475 & FTS & FLOYDS   & 11,500 $\pm$ \phd60 & \phd6.5 $\pm$ 0.7 & \nodata & \nodata & +20.0 \\
60017.172 & MMT & Binospec & 11,950 $\pm$ \phd30 & \phd2.8 $\pm$ 0.8 & \nodata & \nodata & +24.7 \\
60017.514 & FTS & FLOYDS   & 11,720 $\pm$ \phd50 & \phd2.9 $\pm$ 0.5 & \nodata & \nodata & +25.0 \\
\enddata
\tablecomments{The initial spectrum from the Lijiang 2.4\,m Telescope (LJT) was reported to the Transient Name Server by \cite{zhai_lions_2023}. Velocities are derived from the minima of best-fit Gaussians (see Section~\ref{sec:spec}). \\ Only a portion of this table is shown here to demonstrate its form and content. A machine-readable version of the full table is available.}
\end{deluxetable*}\vspace{-24pt}

We obtained optical spectra of SN~2023bee using the Robert Stobie Spectrograph \citep[RSS;][]{smith_prime_2006} on the Southern African Large Telescope \citep[SALT;][]{buckley_status_2006}, the Goodman High-Throughput Spectrograph \citep{clemens_goodman_2004} on the Southern Astrophysical Research Telescope (SOAR), FLOYDS on Las Cumbres Observatory's Faulkes Telescope South \citep[FTS;][]{brown_cumbres_2013}, the Boller \& Chivens (B\&C) Spectrograph on the Bok 2.3\,m Telescope \citep{green_steward_1995}, and Binospec on the MMT \citep{fabricant_binospec:_2019}. We reduced these data using the Image Reduction and Analysis Facility \citep{nationalopticalastronomyobservatories_iraf:_1999}, a custom pipeline based on the PySALT package \citep{crawford_pysalt:_2010}, and the Goodman \citep{torres_goodman_2017}, FLOYDS \citep{valenti_first_2014}, and Binospec \citep{kansky_binospec_2019} pipelines. We also downloaded the classification spectrum of \cite{zhai_lions_2023} from the Transient Name Server and include it in our analysis. The spectra are plotted in Figure~\ref{fig:spec} and logged in Table~\ref{tab:spec}. They are available for download from this Letter and from the Weizmann Interactive Supernova Data Repository \citep{yaron_wiserep_2012}.

\subsection{Extinction and Distance}
We do not detect narrow \ion{Na}{1}~D absorption in any of the spectra of SN~2023bee, implying that dust extinction in the host galaxy is negligible \citep{poznanski_empirical_2012}. Although extinction estimates from low-resolution spectra must be treated with caution \citep{poznanski_lowresolution_2011,phillips_source_2013}, minimal host-galaxy extinction is consistent with the large offset between the SN and the galaxy center and the already blue colors of SN~2023bee. Therefore, we neglect host-galaxy extinction and correct only for Milky Way extinction $E(B-V) = 0.014$~mag \citep{schlafly_measuring_2011} using a \cite{fitzpatrick_correcting_1999} extinction law.

We fit polynomials of order 4--7 to the DLT40 $B$-band photometry of SN~2023bee from $-12$ to $+24$ days of the peak, i.e., excluding the first 4 days of observations. From these polynomials, we find that the light curve peaks on MJD~$59992.3 \pm 0.2$ at $B = 13.24 \pm 0.01$~mag and declines by $\Delta m_{15}(B) = 0.75 \pm 0.03$~mag during the 15 days after peak, where the values and uncertainties are the mean and standard deviation of the values of the four individual fits. According to the \cite{phillips_absolute_1993} relation, we expect an SN~Ia with this slow of a decline to be overluminous, with a peak at $M_B = -19.55 \pm 0.45$~mag \citep{parrent_review_2014}. Including $A_B = 0.06$~mag of extinction, this would imply a distance modulus of $\mu = 32.85 \pm 0.45$~mag or a luminosity distance of $37 \pm 8$~Mpc, which is consistent with estimates of the host-galaxy distance \citep[e.g., $38 \pm 10$~Mpc;][]{springob_erratum_2009} using the method of \cite{tully_new_1977}. We also fit the SALT3 model \citep{kenworthy_salt3_2021} to our multiband DLT40 light curve using SNCosmo \citep{barbary_sncosmo_2022}. Our best-fit parameters ($t_\mathrm{max} = 59993.961 \pm 0.039$, $x_0 = 0.0938 \pm 0.0011$, $x_1 = 1.027 \pm 0.061$, and $c=0.022 \pm 0.012$) indicate an expected distance modulus of 32.45~mag, which is consistent with our previous estimate. Therefore, we adopt $\mu = 32.85$~mag in our analysis and refer to SN~2023bee as an overluminous SN~Ia, keeping in mind that all luminosity-dependent quantities suffer from the aforementioned uncertainty.

\subsection{Radio Nondetections} \label{sec:radioobs}
\begin{figure*}
    \includegraphics[width=0.5\textwidth]{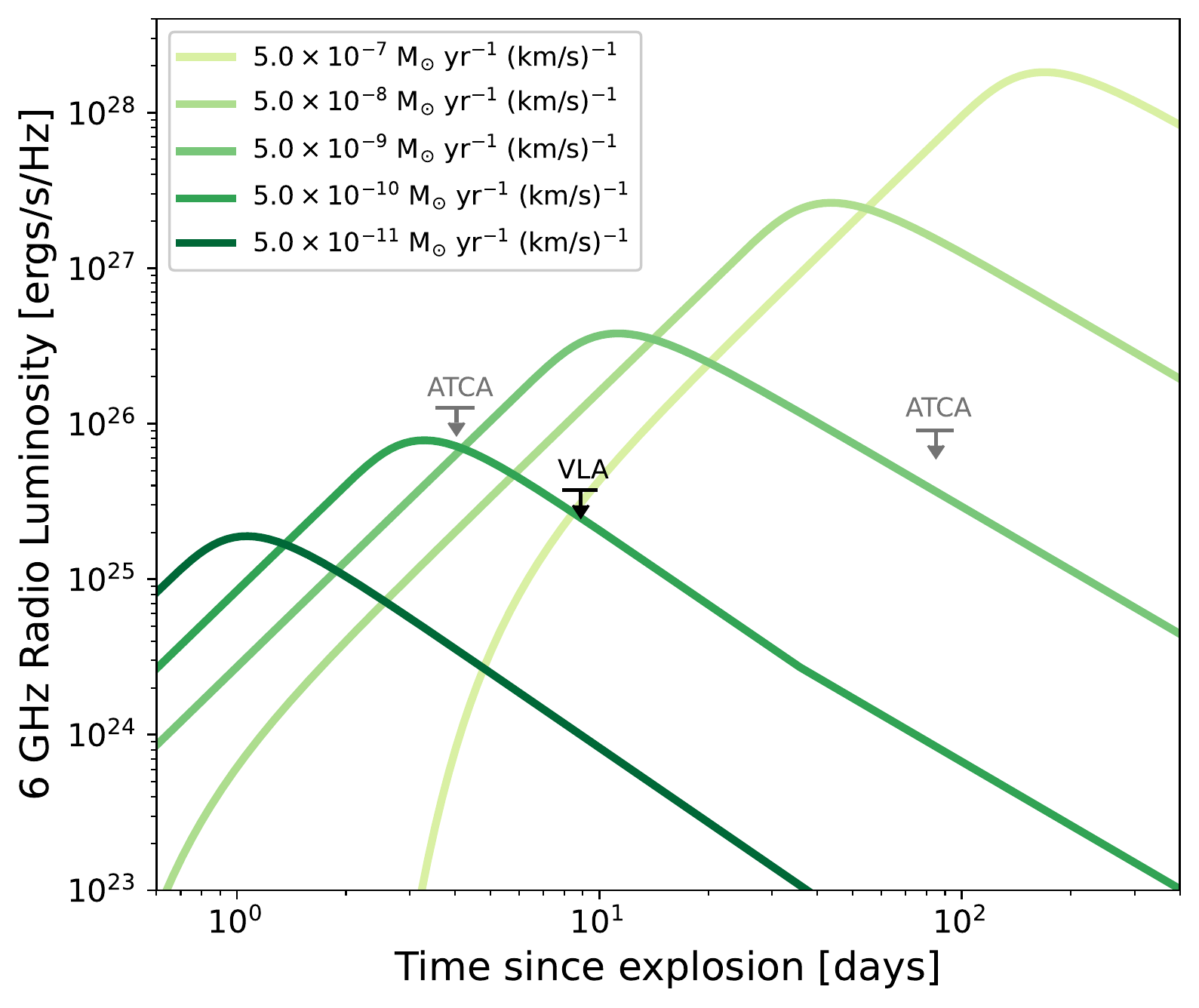}
    \includegraphics[width=0.5\textwidth]{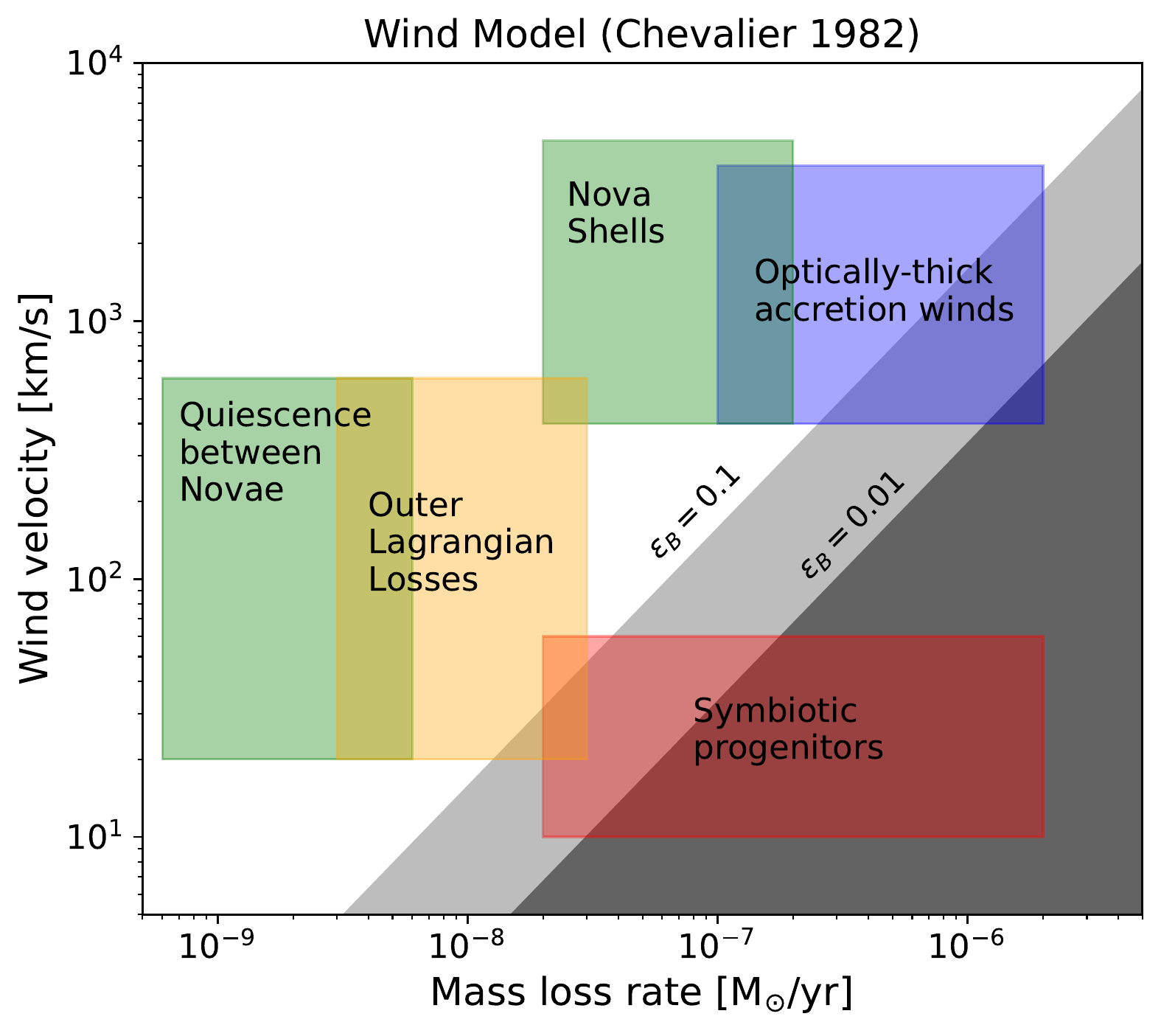}
    \caption{Left: 6~GHz radio luminosity vs.\ time since explosion predicted by model light curves assuming an $r^{-2}$ wind profile with temperature of 10$^5$ K. Values are given for different mass-loss rate/wind velocities ($\dot{M}/v_\mathrm{w}$) assuming $\epsilon_B=\epsilon_e=0.1$. The upper limits correspond to the 5.5~GHz ATCA observations at 4~days from \citet{leung_atca_2023}, as well as at 85 days from the ATCA and at 6~GHz from the VLA, as described in Section~\ref{sec:radioobs}. The 5.5~GHz ATCA observations have been scaled to 6~GHz assuming a spectral index of $-1$.
    Right: parameter space of mass loss from single-degenerate model progenitors as defined by \cite{chomiuk_evla_2012}, such as red-giant mass loss in symbiotic systems, nova eruptions, quiescent mass loss between novae, mass loss through the outer Lagrange points of the white dwarf binary, and fast optically thick winds driven in the high accretion rate regime. The shaded regions are excluded by our radio observations for the case of $\epsilon_B=0.1$ (light shaded) and $\epsilon_B=0.01$ (dark shaded). See Section \ref{sec:radioanalysis} for details.}
    \label{fig:radioobs}
\end{figure*}

We obtained radio observations of SN~2023bee with the Karl G.\ Jansky Very Large Array (VLA) on 2023-02-09.27, $\sim$9 days after our last nondetection. The observation block was 1 hour long, with 38.45 minutes on-source time for SN~2023bee. Observations were taken in $C$-band (4--8~GHz) in the B-configuration of the VLA (DDT: 23A-382, PI: S.\ Sarbadhicary). The observations were obtained in wide-band continuum mode, yielding 4~GHz of bandwidth sampled by 32 spectral windows, each 128 MHz wide sampled by 2 MHz-wide channels with four polarization products. We used 3C138 as our flux, delay, and bandpass calibrator, and J0909+0121 as our complex gain calibrator. 

Data were calibrated with the VLA CASA calibration pipeline (version 6.4.1-12), which iteratively flags corrupted measurements, applies corrections from the online system (e.g. antenna positions), and applies delay, flux density, bandpass and complex gain calibrations \citep{mcmullin_casa_2007}. We then imaged the calibrated visibility dataset with \texttt{tclean} in CASA. We used multiterm, multifrequency synthesis as our deconvolution algorithm (set with \texttt{deconvolver=mtmfs} in \texttt{tclean}), which approximates the full 4--8~GHz wide-band spectral structure of the sky brightness distribution as a Taylor-series expansion about a reference frequency (in our case, 6~GHz) in order to reduce frequency-dependent artifacts during deconvolution \citep{rau_multiscale_2011}. We set \texttt{nterms=2} which uses the first two Taylor terms to create images of intensity (Stokes-I) and spectral index. We sampled the synthesized beam with 0\farcs3~pixels and imaged out to 10\farcm24 to deconvolve any outlying bright sources and mitigate their sidelobes at the primary beam center. Gridding was carried out with the W-projection algorithm (\texttt{gridder=wproject}) with 16 w-planes. Images were weighted with the Briggs weighting scheme (\texttt{weighting=briggs}) using a robust value of 0 to balance point-source sensitivity with high angular resolution and low sidelobe contamination between sources. The final image has a spatial resolution of $1\farcs02 \times 0\farcs85$ (or $183\times153$ pc, assuming a distance of 37~Mpc), and an RMS of about 4.5~$\mu$Jy~beam$^{-1}$.

No radio source was detected at the site of SN~2023bee in the cleaned, deconvolved 6~GHz image at the $3\sigma$ level. The flux at the SN location is $8.3~\mu$Jy~beam$^{-1}$, and the RMS noise in a circular region $\sim$28\arcsec\ across, centered on the SN, is $4.8~\mu$Jy~beam$^{-1}$. We therefore assume a flux density upper limit of 22.7 $\mu$Jy, which is equal to the flux density at the SN location plus $3\sigma$ noise to be consistent with previous work \citep{chomiuk_evla_2012}. At a distance of 37~Mpc, this corresponds to an upper limit on 6~GHz luminosity of $3.7 \times 10^{25}$
ergs s$^{-1}$ Hz$^{-1}$ (Figure~\ref{fig:radioobs}). 

In addition to the VLA, we also made use of radio observations from the Australia Telescope Compact Array (ATCA). \citet{leung_atca_2023} obtained 3$\sigma$ upper limits of 84~$\mu$Jy at 5.5~GHz, 60~$\mu$Jy at 9.0 GHz, 87~$\mu$Jy at 16.7~GHz, and 129~$\mu$Jy at 21.2 GHz within 5 days of the explosion. We also obtained new observations at 5.5~GHz and 9.0~GHz at 85 days past explosion to look for any late-time emission due to denser winds. These data were processed in the manner described by \citet{bufano_sn_2014}, and yielded $3\sigma$ upper limits of 60 and 30~$\mu$Jy at 5.5 and 9.0~GHz respectively. The 5.5 GHz ATCA limits are also plotted in Figure~\ref{fig:radioobs}, yielding an upper limit on 6 GHz luminosity of $9 \times 10^{25}$~ergs~s$^{-1}$~Hz$^{-1}$.

\section{Light Curves and Color Curves} \label{sec:phot}
\begin{figure*}
    \centering
    \includegraphics[width=0.85\textwidth]{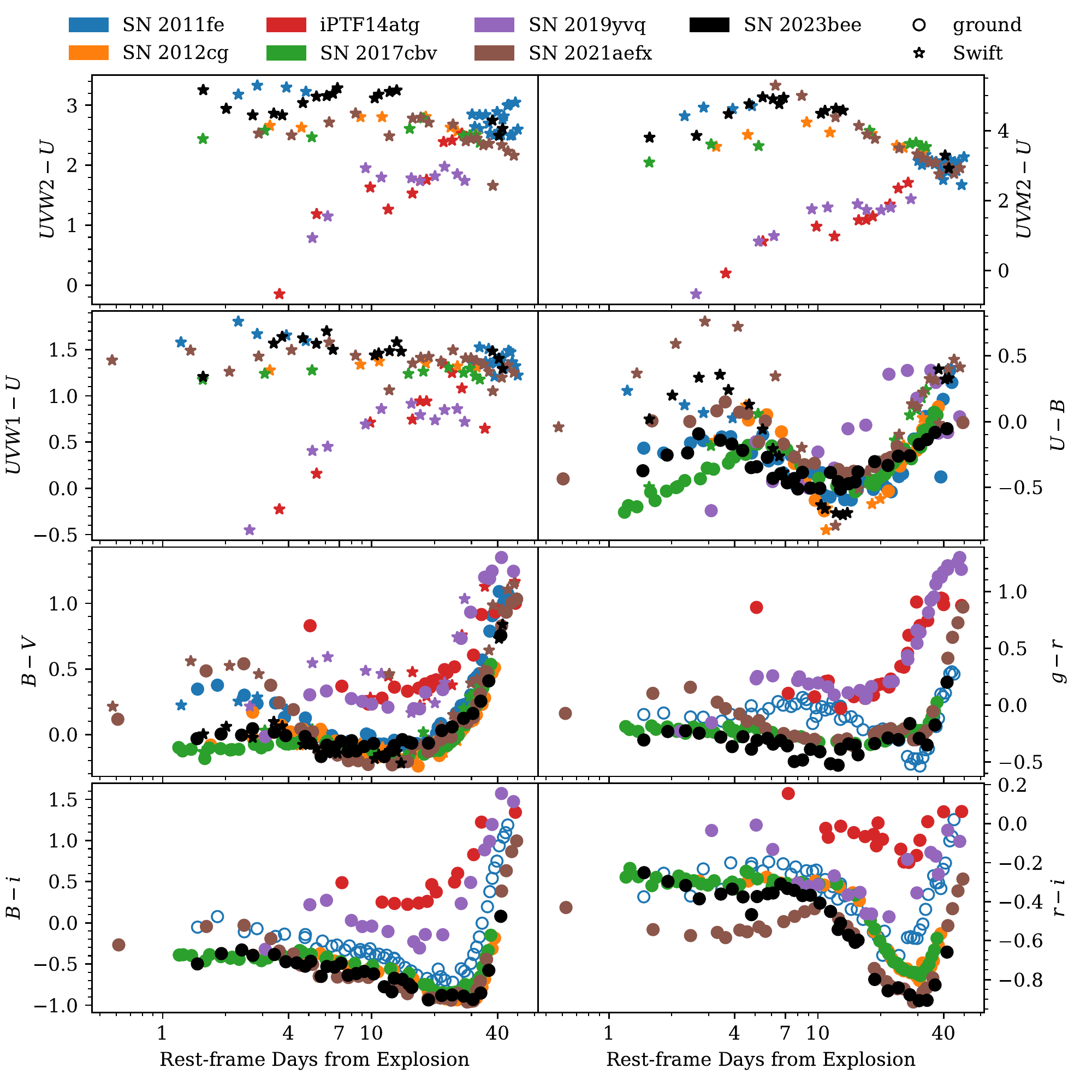}
    \caption{Color curves of SN~2023bee (plotted in black), corrected for Milky Way extinction, compared to other Type~Ia supernovae: SN~2011fe \citep{zhang_optical_2016}, which did not show an early excess; SNe~2012cg \citep{marion_sn_2016}, 2017cbv \citep{hosseinzadeh_early_2017}, and 2021aefx \citep{hosseinzadeh_constraining_2022}, which were normal to overluminous and showed an early excess; and iPTF14atg \citep{cao_strong_2015} and SN~2019yvq \citep{burke_bright_2021}, which were underluminous and showed an early excess. Open circles for SN~2011fe are converted from \textit{VRI} to \textit{gri} using the relationships of \cite{jordi_empirical_2006}. All the normal to overluminous events with an early excess show similar color behavior during the first several days after explosion: a rapid reddening prior to landing near the color curve of SN~2011fe. This reddening phase corresponds to the excess in $U$ and other bands. In the \cite{kasen_seeing_2010} model, in which this excess is caused by a collision with a binary companion, the duration of this phase is proportional to the binary separation. \\ (The data used to create this figure are available.)}
    \label{fig:colors}
\end{figure*}

Figure~\ref{fig:colors} shows the color curves of SN~2023bee compared to other SNe~Ia. Like other overluminous SNe with an early $U$-band excess, namely SNe~2017cbv and 2021aefx, there is a sharp change in slope in the $U-B$ color, and to a lesser extent $B-V$, in the first few days after explosion. This indicates an initially very blue color, rapidly reddening until it matches the slope of the normal SN~Ia color curve (e.g., that of SN~2011fe). In SN~2023bee, this initial reddening phase lasts $\sim$3 days, the same phase that the excess is visible in the $U$-band light curve. This indicates an additional blue component not present in all SNe~Ia that dominates the emission during the first few days.

\begin{deluxetable*}{lLlllLl}
\tablecaption{Companion-shocking Model Parameters\label{tab:params}}
\tablehead{\colhead{Parameter} & \colhead{Variable} & \colhead{Prior Shape} & \twocolhead{Prior Parameters\tablenotemark{a}} & \colhead{Best-fit Value\tablenotemark{b}} & \colhead{Units}}
\tablecolumns{7}
\startdata
\cutinhead{Companion-shocking Model \citep{kasen_seeing_2010}}
Explosion time & t_0 & Uniform & 59975.12 & 59975.65 & 59975.3^{+0.2}_{-0.1} & MJD \\
Binary separation & a & Uniform & 0 & 1 & 0.055^{+0.010}_{-0.009} & $10^{13}$ cm \\
Viewing angle & \theta & Uniform & 0 & 180 & 0 \pm 16 & deg \\
\cutinhead{SiFTO Model \citep{conley_sifto:_2008}}
Time of $B$ maximum & t_\mathrm{max} & Uniform & 59991.50 & 59993.50 & 59992.04 \pm 0.03 & MJD \\
Stretch & s & Log-uniform & 0.5 & 2 & 0.988 \pm 0.004 & dimensionless \\
Time shift in $U$ & \Delta t_U & Gaussian & 0 & 1 & +0.09 \pm 0.05 & d \\
Time shift in $i$ & \Delta t_i & Gaussian & 0 & 1 & +0.29 \pm 0.05 & d \\
\cutinhead{Combined Model}
Intrinsic scatter & \sigma & Half-Gaussian & 0 & 1 & 11.0 \pm 0.3 & dimensionless
\enddata
\tablenotetext{a}{The ``Prior Parameters'' columns list the minimum and maximum for a uniform distribution, and the mean and standard deviation for a Gaussian distribution.}
\tablenotetext{b}{The ``Best-fit Value'' column is determined from the 16th, 50th, and 84th percentiles of the posterior distribution, i.e., $\mathrm{median} \pm 1\sigma$, with one exception. The posterior for the viewing angle is approximately half-Gaussian peaking at zero, so we use the 0th and 68th percentiles.}
\end{deluxetable*}\vspace{-24pt}

\begin{figure*}
    \includegraphics[width=0.5\textwidth]{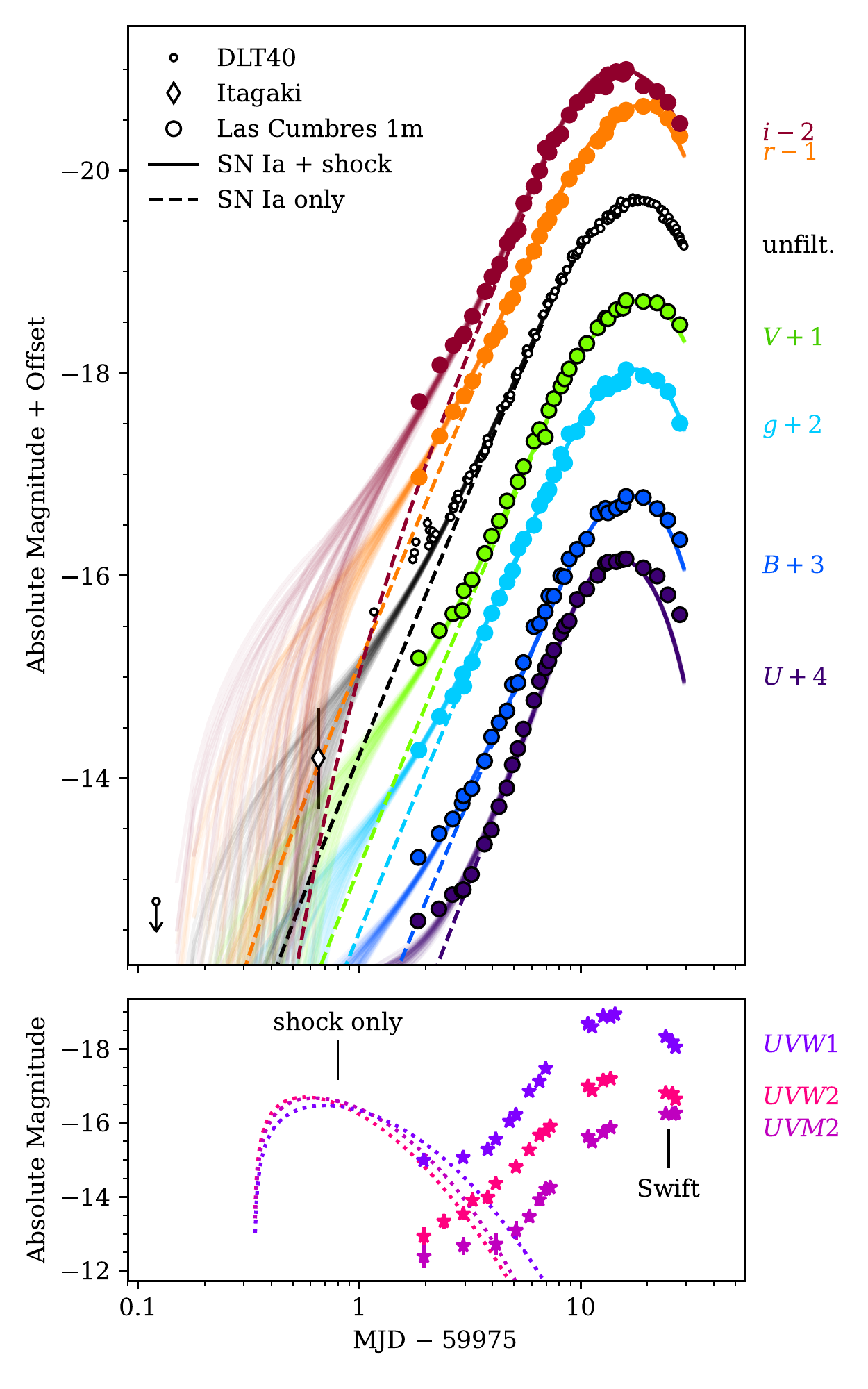}
    \includegraphics[width=0.5\textwidth]{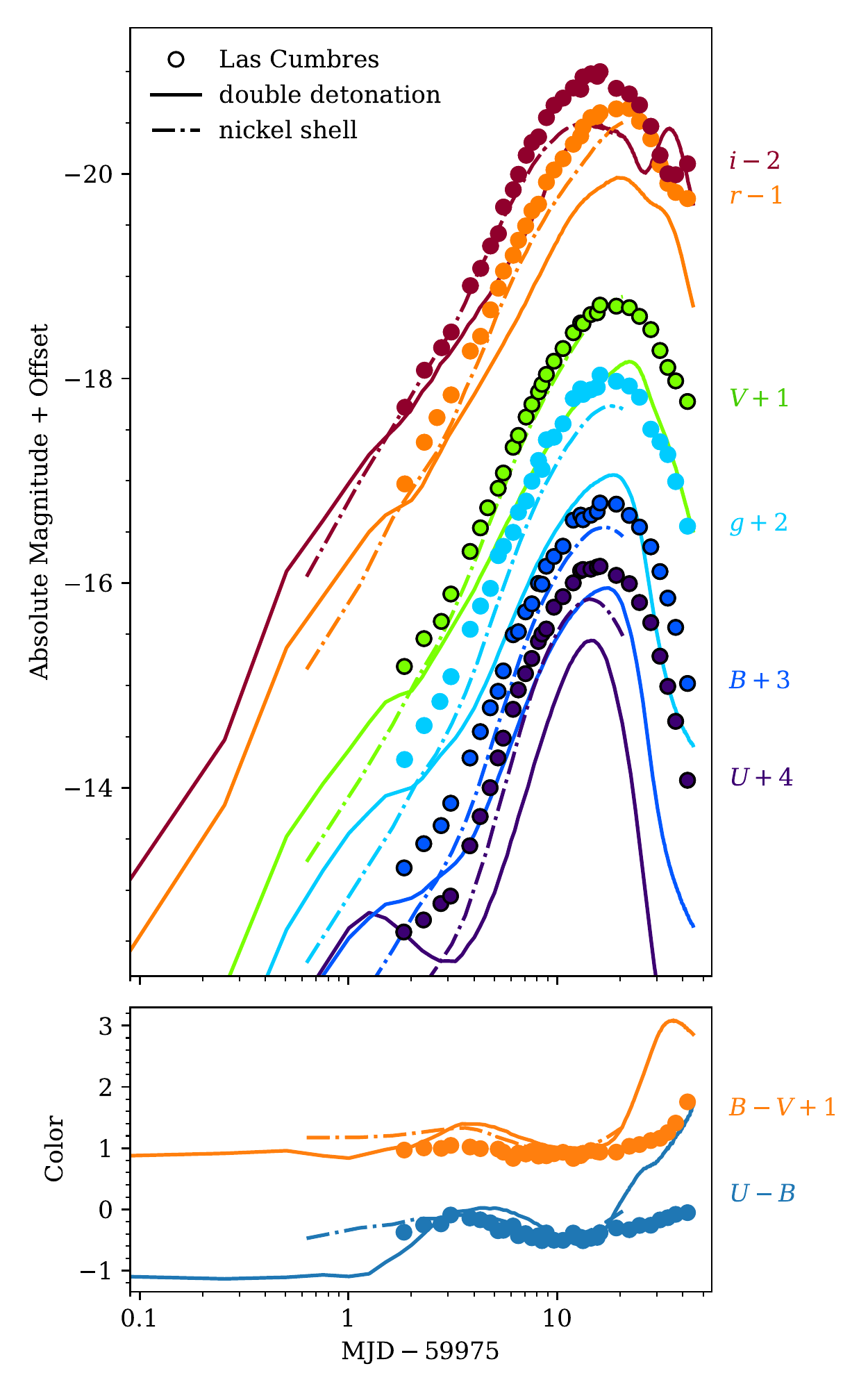}
    \caption{Top left: the light curve of SN~2023bee compared to the model of \cite{kasen_seeing_2010}, in which the SN ejecta collide with a nondegenerate binary companion. The solid lines are random draws from the posterior, summarized in Table~\ref{tab:params}. The dashed lines show the ``normal'' Type~Ia component of this model, the SiFTO template from \cite{conley_sifto:_2008}. All bands show an excess with respect to this template for the first $\sim$4 days, with stronger excesses in bluer bands.
    Bottom left: the shock component of the model alone (SiFTO does not extend into the UV) compared to the Swift photometry of SN~2023bee. The model correctly predicts the \textit{UVW1} band, unlike in previous events in the literature, but greatly overpredicts \textit{UVM2} and \textit{UVW2}, likely because the blackbody assumption is not valid.
    Top right: the light curve of SN~2023bee compared to the best-matched double-detonation model of \cite{polin_observational_2019} and the best-matched nickel-shell model of \cite{magee_investigation_2020}. Neither provides a good fit to the data both at peak and during the first few days.
    Bottom right: the color curves of SN~2023bee compared to the same two models above. The best-matched models were chosen based on the lowest $\bar\chi^2$ for the color curves. Although the color curves are a reasonable match during the first few days, the light curves (above) are not.}
    \label{fig:fit}
\end{figure*}

To investigate the origin of this blue component, we compare the multiband light curve to several models from the literature. First, we consider the model of \cite{kasen_seeing_2010} in which the ejecta collide with and are shocked by a nondegenerate binary companion. In this model, the strength and duration of the excess depend mostly on the binary separation $a$, or equivalently the companion radius $R$, where $a/R = 2{-}3$ if we assume it is in Roche-lobe overflow. We construct a model consisting of the SiFTO SN~Ia template from \cite{conley_sifto:_2008} plus an analytical shock component as described by \cite{kasen_seeing_2010}, where the latter is suppressed based on the viewing angle according to the approximation of \cite{brown_constraints_2012}. (See \citealt{olling_no_2015} for an alternative approximation.) In this model, the luminosity scale of the shock component is proportional to the product $a (M v^7)^{1/4} f(\theta)$, where $M$ is the ejecta mass, $v$ is the transition velocity between density profiles in the outer ejecta, and $f(\theta)$ is the fraction of the flux that reaches the observer as a function of the viewing angle. The temperature, and therefore color, of the shock emission also depends on $a$, so it can be fit for separately. However, it is not practical to constrain the remaining parameters simultaneously. Therefore we fix $Mv^7 = M_\mathrm{Ch} (10{,}000\mathrm{\ km\ s}^{-1})^7$, i.e., a Chandrasekhar mass of ejecta, the outer layers of which are moving at 10,000~km~s$^{-1}$. We caution that fixing this parameter, which can vary by orders of magnitude due to the large exponent, might cause us to underestimate uncertainties on the other parameters. Off-axis systems with high ejecta masses or velocities look the same as on-axis systems with low ejecta masses or velocities.

Because our focus is on the shock component and not on the physics powering the light-curve peak, we scale each band of the SiFTO model so that it matches the observed maximum of SN~2023bee without preserving the colors of the original template. We also fit for small time shifts in $U$ and $i$ relative to the other bands in order to better match the observed light curve. We only include the first 30 days of data in the fit (until MJD 60005), because the SiFTO template does not match our data well after this time, specifically the slope of the $U$-band decline and the secondary infrared maximum. The SiFTO template does not extend into the ultraviolet, so we cannot include our Swift data in the fit. For efficiency, we also exclude the filtered DLT40 photometry, and we bin the remaining points by $\sim$0.01 days. Finally, we include an intrinsic scatter term $\sigma$ that effectively inflates the photometric uncertainties by a factor of $\sqrt{1+\sigma^2}$. We perform the fit using an MCMC routine implemented in the Light Curve Fitting package \citep{hosseinzadeh_light_2023}. Table~\ref{tab:params} describes the prior and posterior distributions of the fit parameters, and Figure~\ref{fig:fit} (left) shows the best-fit model compared to our observed light curve.

The model provides an excellent fit to the excess in the $U$ band, and it reveals weaker excesses in all other filters compared to the SiFTO template. The best-fit binary separation implies a companion radius of a few solar radii, i.e., a main-sequence star. The posterior for the viewing angle peaks at zero, with a 68th percentile at $16\degr$, i.e., $\theta = 0\ \pm 16\degr$. Note that the posterior for the binary separation covaries weakly with the posterior for the viewing angle because a larger companion off-axis has the same effect as a smaller companion on-axis. As shown in the bottom left panel of Figure~\ref{fig:fit}, the shock component alone overpredicts the Swift photometry by $\sim$2~mag in \textit{UVM2} and \textit{UVW2}. This effect was also seen in SNe~2017cbv and 2021aefx and can be attributed to the assumption that the shock component radiates as a blackbody \citep{hosseinzadeh_early_2017,hosseinzadeh_constraining_2022}. In reality, the shock radiation may be subject to line blanketing when it is reprocessed by the SN ejecta, suppressing the UV flux. Unlike in previous events, the shock model is a reasonable fit to the \textit{UVW1} (longest-wavelength UV) photometry, which also shows a change in slope between the second and third observations.

To confirm the excesses at other wavelengths, we repeat our fitting procedure after excluding the $U$-band data. We find that the posteriors do not change significantly. While there is no requirement that the same mechanism produces the excesses in all bands, a complete explanation requires an effect at all optical to near-UV wavelengths, with a larger effect in the bluer bands. As the early spectra do not display unusual features (see Figure~\ref{fig:spec}), the simplest explanation is an additional hot thermal component to the emission, possibly reprocessed through the SN ejecta. We discuss this further in Section~\ref{sec:discuss}.

We also consider the double-detonation models of \cite{polin_observational_2019}, in which a shell of surface helium detonates before igniting the core of the white dwarf. Some of these models show an initial blue bump in their light curves. We calculated the $U-B$ and $B-V$ colors of the models and compared them to our observations, applying time offsets from 0 to 3 days at 0.5 day intervals and calculating the reduced $\chi^2$ ($\bar\chi^2 \equiv \chi^2/N$, where $N$ is the number of observed points that overlap the model in time) for each model in the grid. The best-matched model is of a $0.9\ M_\sun$ white dwarf edge-lit by a $0.09\ M_\sun$ helium shell, offset by 0.5 days, with $\bar\chi^2=3342$ and $N=32$. This model is shown in the right panel of Figure~\ref{fig:fit}. While the model does reasonably reproduce the early color curve of SN~2023bee, both the $U$-band and $B$-band excesses are much stronger than observed. In addition, the peak of the light curve is a very poor match. This is expected: SN~2023bee is overluminous, whereas the double-detonation models of \cite{polin_observational_2019} are underluminous (though see \citealt{shen_multidimensional_2021} for overluminous double detonations).

Another model for early light-curve excesses comes from shells or clumps of radioactive $^{56}$Ni near the surface of the ejecta, providing an extra source of photons that can quickly escape to the observer. \cite{magee_investigation_2020} produce a set of such models designed to match SNe~2017cbv and 2018oh, both of which had early light-curve excesses. We compare these models to our observed color curves using the same prodecure as above and find that the best-matched model requires a shell of $0.02\ M_\sun$ of $^{56}$Ni centered at $1.35\ M_\sun$ with a thickness of $0.18\ M_\sun$ in mass coordinates, offset by 0.5 days, which has $\bar\chi^2=116$ with $N=25$. Again we see a reasonable match to the observed $U-B$ and $B-V$ color curves, but this time the bump is not strong enough to match the observed light curve (Figure~\ref{fig:fit}, right).

For both the double-detionation and nickel-shell models, we were not able to find a better match by comparing the light curves directly, rather than the color curves. There are better matches to the light-curve peaks, but these do not correctly predict the early excess. Comparing these models to SN~2021aefx yielded the same problem \citep{hosseinzadeh_constraining_2022}.

Finally, we compared models of interaction with CSM from \cite{piro_exploring_2016} and models of pulsational delayed detonations from \cite{dessart_constraints_2014} to our observed light curves and color curves using the procedure above. However, the models in these sets are significantly poorer matches to our data than the three models discussed previously, so we do not discuss them further.

\section{Early Spectra} \label{sec:spec}
The features in our earliest spectra indicate very high velocities for the material ejected in the explosion. For example, the blue edge of the \ion{Si}{2} line at 636~nm in our first spectrum implies a maximum velocity of $\sim$42,000~km~s$^{-1}$. The asymmetry of this feature in spectra more than 1 week before maximum light likely indicates two or more velocity components, as well as blending with the \ion{Si}{2} line at 597~nm. The broad, blended, hydrogen-free features of very young SNe~Ia, in combination with the scarcity of equally early spectra to compare to in the literature, have led to mistaken Type~Ic or uncertain Type~I classifications in this event and others \citep{bostroem_global_2021,zhai_lions_2023}.

We measured the expansion velocity and equivalent width of the \ion{Si}{2} line at 636~nm in each of our spectra by fitting a single Gaussian line profile in absorption on a linear continuum using a Markov Chain Monte Carlo (MCMC) routine implemented in the \texttt{emcee} package \citep{foreman-mackey_emcee_2013}. This is not always a perfect representation of the complete line profile, but it is sufficient to measure the velocity of the absorption minimum. In cases where \ion{C}{2} is also visible at 658~nm, we simultaneously fit a second Gaussian in absorption. The velocities are derived from the center of the Gaussian component of the model, and the equivalent widths come from integrating the data, after subtracting the model of the other line when necessary, divided by the model continuum. These quantities are also reported in Table~\ref{tab:spec}.

Absorption due to \ion{C}{2} is clearly visible in our first five spectra ($>$14 days before peak; $<$3 days after explosion), with photospheric velocities around 21,000~km~s$^{-1}$ and equivalent widths around 3.3~nm. Carbon is a signature of unburned material, which is not predicted by all explosion models, and has been previously observed in a significant fraction of early SN~Ia spectra (e.g., \citealt{parrent_study_2011,thomas_type_2011,folatelli_unburned_2012,silverman_berkeley_2012a,maguire_exploring_2014}).

Another remarkable feature of our measurements is that the equivalent width of \ion{Si}{2} initially increases from $9.8 \pm 0.9$~nm in our first spectrum to $14.1 \pm 1.2$~nm a few days later, followed by a decline until around maximum light. The shallow absorption lines in our earliest spectra suggest dilution of the lines by an additional continuum component of the emission. \cite{marion_sn_2016} similarly noted shallow spectral features in spectra of SN~2012cg, which also showed an early light-curve excess, at these same phases. It is unclear whether this is normal behavior; for example, the earliest equivalent width measurements in the sample of \cite{silverman_berkeley_2012a} are $\sim$13 days before maximum light. The extra continuum could come from several of the processes we discuss above: collision with a binary companion, extra heating from surface nickel, or interaction with CSM.

\cite{ashall_speed_2022} postulated that the early $U$/$u$-band excess in SN~2021aefx was caused purely by the strong \ion{Ca}{2} H and K absorption feature traversing the $U$/$u$ transmission function as the photosphere decelerates with time. However, the earliest observed velocity in SN~2023bee is $\sim$25,000~km~s$^{-1}$, which is near the minimum requirement for a $U$-band bump according to \cite{ashall_speed_2022}. This mechanism is not able to explain excesses in most other bands, including the clear deviation of the unfiltered light curve from a power law, as well as the weaker excesses in all optical bands implied by our fitting.

To demonstrate the effect of SN~2023bee's lower early velocity on the early $U-B$ color in the ``speed bump'' paradigm of \cite{ashall_speed_2022}, we blueshift our SALT spectrum taken 15.2 days before peak (the first-night spectrum that extends farthest into the blue) from a photopheric velocity of 24,000 to 30,000~km~s$^{-1}$ to simulate the earliest velocity of SN~2021aefx. We then measure synthetic photometry in $U$ and $B$ on the original and blueshifted spectra. In $U$, where our spectrum does not cover the full passband, we only integrate over the wavelengths covered by the spectrum. We find that the higher-velocity spectrum is bluer by $\Delta(U-B) = -0.2$~mag, which is approximately how much bluer SN~2021aefx is than SN~2023bee. Considered alongside the evidence presented above, we conclude that, while this velocity effect may cause SNe~Ia to be bluer during the first few days after explosion, particularly in the $U$ and $u$ bands, it cannot be the sole explanation for the early light-curve behavior of either of these events.

\section{Radio Constraints} \label{sec:radioanalysis}
We can independently assess possible single-degenerate progenitors by exploring what properties of wind-like CSM are allowed by our radio upper limits (Section~\ref{sec:radioobs}). Following the methodology of \cite{chomiuk_evla_2012,chomiuk_deep_2016}, we assume SN~2023bee was surrounded by the \cite{chevalier_radio_1982} model of CSM, produced by steady mass loss from the progenitor, i.e., $\rho_\mathrm{csm} = \dot{M}/4 \pi r^2 v_\mathrm{w}$, where $\rho_\mathrm{csm}$ is the CSM density, $\dot{M}$ is the mass-loss rate from the progenitor, $r$ is the distance from the progenitor, and $v_\mathrm{w}$ is the wind velocity. We fix the energy fraction in relativistic electrons to $\epsilon_e = 0.1$. We note that \cite{chomiuk_evla_2012} assumes synchrotron self-absorption at low frequencies, but at early times such as our VLA observation, free-free absorption can also become important for high-density winds. We modify the optically thin luminosity from \cite{chomiuk_evla_2012} by a factor $\mathrm{e}^{-\tau_{\mathrm{ff}}}$ (where $\tau_{\mathrm{ff}}$ is the free-free optical depth) to account for free-free absorption. We use the definition of $\tau_{\mathrm{ff}}$ given in \cite{fransson_circumstellar_1996} as 
\begin{equation} \label{eq:tauff}
\tau_{\mathrm{ff}} = 17.8\ \nu_{6}^{-2} \dot{M}_{-5}^{2} v_{\mathrm{w},1}^{-2} T_{5}^{-3/2} (v_{4}t_{10})^{-3}
\end{equation}
where $\dot{M}_{-5} = \dot{M}/10^{-5}\ M_\sun\ \mathrm{yr}^{-1}$ is the mass-loss rate, $v_{\mathrm{w},1} = v_\mathrm{w}/10$ km s$^{-1}$ is the wind speed, $T_{5}=T/10^5$ K is the wind temperature, $v_{4}=v/10^4\mathrm{\ km\ s}^{-1}$ is the ejecta velocity, $t_{10}=t/10$ days is the SN age, and $\nu_6=\nu/6$~GHz is our observing frequency. We assume $T_5=1$ for our calculations.\footnote{Typically the wind temperature is expected to vary between $T_5 = 0.2{-}1$ \citep{lundqvist_hydrogen_2013}, but this primarily affects the rising part of the light curve and hence the upper limit to the mass-loss rate discussed later. For $T_5=0.2$, we obtain an upper limit of $10^{-3}\ M_\sun\ \mathrm{yr}^{-1}$ for $v_\mathrm{w}=100$ km~s$^{-1}$ from the 9~GHz upper limit, which still excludes red giant winds.} The resulting light curves for different mass-loss rates are shown in Figure \ref{fig:radioobs}. 

Assuming a standard SN Ia explosion with $10^{51}$~erg kinetic energy and $1.4\ M_\sun$ ejecta mass, we obtain a mass-loss rate upper limit of $\dot{M}/v_\mathrm{w} < (6.4{-}30.5) \times 10^{-10}\ M_\sun\mathrm{\ yr^{-1}\ km^{-1}\ s}$. The range of mass-loss rates reflects the uncertainty in the parameter $\epsilon_B$, the fraction of shock energy shared by the amplified magnetic field, with typical values in the range 0.01--0.1 for SNe \citep{chomiuk_evla_2012}. These limits are also consistent with the reported upper limits from ATCA \citep{leung_atca_2023}, particularly the 5.5 GHz (shown in Figure \ref{fig:radioobs}) and 9 GHz limits,\footnote{For clarity we only show the scaled 5.5~GHz and 6~GHz data, at which we have the most observations. The other frequencies (9, 16.7, and 21~GHz) provide similar or slightly shallower constraints.} which, though shallower than our 6~GHz VLA data, are similarly constraining, because they were taken a few days earlier when the light curve would have been peaking. 

These limits are compared with the mass-loss-rate parameter space of single-degenerate models, as defined by \cite{chomiuk_evla_2012}, in Figure \ref{fig:radioobs}. We find that our limits are deep enough to rule out the majority of red-giant companions (symbiotic systems), characterized by slow winds of 10--100~km~s$^{-1}$ and mass-loss rates of $10^{-6}{-}10^{-8}\ M_{\odot}\ \mathrm{yr}^{-1}$ \citep{seaquist_collective_1990}. Symbiotic systems have also been ruled out for the majority of SNe Ia based on radio upper limits \citep{chomiuk_evla_2012,horesh_early_2012,perez-torres_constraints_2014,chomiuk_deep_2016,lundqvist_deepest_2020,pellegrino_constraining_2020,burke_bright_2021}. Many models involving main-sequence companions for SN~2023bee, however, are still allowed within our limits (see the colored regions in Figure~\ref{fig:radioobs}). Weak constraints are obtained for the channel of fast, optically thick winds from white dwarfs accreting at high rates, while more quiescent mass-loss and nova-like densities are still allowed by our observations. 

We briefly note that, for mass-loss rates ${>}3 \times 10^{-3}\ M_\sun\ \mathrm{yr}^{-1}$ (assuming $v_\mathrm{w}=100$~km~s$^{-1}$), free-free absorption of radio emission will still yield nondetection at our 85~day ATCA observations at 5.5 and 9~GHz (for the parameters in Eq.~\ref{eq:tauff}). However, such high mass-loss rates in thermonuclear SNe have only been observed in a rare variety of Type Ia CSM, which shows signatures of dense CSM interaction, such as broadened emission lines of hydrogen or helium in the spectra \citep{silverman_type_2013}. No such features are seen in the spectra of SN~2023bee. 

Our radio upper limits therefore make it unlikely that a red-giant companion was part of SN~2023bee's progenitor system, unless the ratios of energies in the magnetic fields and relativistic electrons to the shock energy are much smaller than typically assumed (i.e., $\epsilon_B, \epsilon_e \ll 1\%{-}10\%$). A red-giant companion is also effectively ruled out by our light-curve fitting in Section~\ref{sec:phot}. However, the main-sequence companion preferred by our light-curve fitting is still permitted by our radio upper limits.

\section{Discussion} \label{sec:discuss}
Taken together, our observations above suggest that the emission from SN~2023bee during the first few days after explosion consists of typical SN~Ia spectral features plus an additional hot continuum component. This causes an excess in all bands but preferentially at blue to UV wavelengths. Of the models we consider, the companion-shocking model of \cite{kasen_seeing_2010} best reproduces the strength, duration, and color of the excess in the \textit{UVW1} through the $i$ bands. Other models, including those involving a double detonation or a shell of $^{56}$Ni near the surface of the ejecta, are able to produce an excess of approximately the same duration but do not match in either strength or color. The worse match of these latter models might be attributed to their relatively small grid of parameters, while the analytic companion-shocking model is more flexible. Larger model grids or analytic approximations could allow for fairer comparisons in the future.

One weakness of the \cite{kasen_seeing_2010} model, already noted in previous work \citep{hosseinzadeh_early_2017,hosseinzadeh_constraining_2022}, is the assumption that the shock component of the emission does not suffer from the same line blanketing in the UV as the emission from the rest of the ejecta. This leads to an overprediction in the farther-UV bands (\textit{UVM2} and \textit{UVW2}). The severity of this overpredicton likely depends on the details of the spectrum, including line strengths and velocities, which in turn depend on the conditions in the ejecta and possibly the viewing angle. Future modeling work should focus on a more realistic spectral energy distribution for the shock component with the goal of a fairer comparison with the data.

In this single-degenerate scenario, the ejecta are expected to strip material off the companion star during the collision, and this material could emit at late phases when the ejecta are mostly transparent \citep{botyanszki_multidimensional_2018,dessart_spectral_2020}. SN~2023bee has not yet reached this phase, but the search for nebular emission will be an important test of the single-degenerate model. Previous searches for hydrogen and helium in nebular spectra of SNe~Ia have mostly yielded null results (\citealt{leonard_constraining_2007,graham_nebular-phase_2017,sand_nebular_2019,tucker_nebular_2020}), even in cases where there was an early light-curve excess (\citealt{sand_nebular_2018,dimitriadis_nebular_2019,tucker_no_2019,siebert_strong_2020,tucker_sn_2021,hosseinzadeh_constraining_2022}). However, translating line luminosity limits to limits on the stripped mass depends on complex and poorly understood physical processes.

This Letter demonstrates the power of using very high-cadence, multiband photometry of young, nearby SNe~Ia to constrain their progenitor systems, which is only possible with specially designed robotic facilities like the DLT40 Survey, Las Cumbres Observatory, and Swift. As the number of well-observed events grows, increasingly high-quality data sets are beginning to reveal the weaknesses in existing models of SN~Ia progenitors and explosion physics. Close cooperation between observers and modelers is required to narrow this gap.

\section*{Acknowledgments}
This work makes use of data from the Las Cumbres Observatory telescope network.
The SALT data presented here were taken via the Rutgers University program 2022-1-MLT-004 (PI: S.W.J.) and the South African program 2021-2-LSP-001 (PI: D.A.H.B.). Polish participation in SALT is funded by grant No.\ MEiN nr2021/WK/01.
The Australia Telescope Compact Array is part of the Australia Telescope National Facility (\url{https://ror.org/05qajvd42}) which is funded by the Australian Government for operation as a National Facility managed by CSIRO. We acknowledge the Gomeroi people as the traditional owners of the Observatory site.
The National Radio Astronomy Observatory is a facility of the National Science Foundation operated under cooperative agreement by Associated Universities, Inc.

Time domain research by G.H, D.J.S, and the University of Arizona team is supported by NSF grants AST-1821987, 1813466, 1908972, and 2108032, and by the Heising-Simons Foundation under grant \#2020--1864. 
J.E.A.\ is supported by the international Gemini Observatory, a program of NSF's NOIRLab, which is managed by the Association of Universities for Research in Astronomy (AURA) under a cooperative agreement with the National Science Foundation, on behalf of the Gemini partnership of Argentina, Brazil, Canada, Chile, the Republic of Korea, and the United States of America.
This publication was made possible through the support of an LSSTC Catalyst Fellowship to K.A.B., funded through Grant 62192 from the John Templeton Foundation to LSST Corporation. The opinions expressed in this publication are those of the authors and do not necessarily reflect the views of LSSTC or the John Templeton Foundation.
Research by S.V., Y.D., E.H., and N.M.\ is supported by NSF grant AST-2008108.
The Las Cumbres Observatory group is supported by NSF grants AST-1911225 and 1911151.
L.A.K.\ acknowledges support by NASA FINESST fellowship 80NSSC22K1599.
D.A.H.B.\ acknowledges research support from the National Research Foundation of South Africa.
L.C.\ is grateful for support from NSF grant AST-2107070.

\facilities{ADS, ATCA, Bok (B\&C), CTIO:PROMPT, LCOGT (Sinistro, FLOYDS), Meckering:PROMPT, MMT (Binospec), NED, SALT (RSS), SOAR (Goodman), Swift (UVOT), VLA}

\defcitealias{astropycollaboration_astropy_2022}{Astropy Collaboration 2022}
\software{Astrometry.net \citep{lang_astrometry_2010}, Astropy \citepalias{astropycollaboration_astropy_2022}, BANZAI \citep{mccully_lcogt_2018}, CASA \citep{mcmullin_casa_2007}, \texttt{corner} \citep{foreman-mackey_corner.py:_2016}, \texttt{emcee} \citep{foreman-mackey_emcee_2013}, extinction \citep{barbary_extinction_2016}, FLOYDS pipeline \citep{valenti_first_2014}, Goodman pipeline \citep{torres_goodman_2017}, HEAsoft \citep{nasaheasarc_heasoft:_2014}, IRAF \citep{nationalopticalastronomyobservatories_iraf:_1999}, \texttt{lcogtsnpipe} \citep{valenti_diversity_2016}, Light Curve Fitting \citep{hosseinzadeh_light_2023}, Matplotlib \citep{hunter_matplotlib:_2007}, NumPy \citep{oliphant_guide_2006}, Photutils \citep{bradley_astropy_2022}, PyRAF \citep{sciencesoftwarebranchatstsci_pyraf:_2012}, PySALT \citep{crawford_pysalt:_2010}, \texttt{reproject} \citep{robitaille_image_2023}, SNCosmo \citep{barbary_sncosmo_2022}}

\bibliography{zotero_abbrev}

\end{document}

%% file: affil.tex
\newcommand{\LCO}{\affiliation{Las Cumbres Observatory, 6740 Cortona Drive, Suite 102, Goleta, CA 93117-5575, USA}}
\newcommand{\UCSB}{\affiliation{Department of Physics, University of California, Santa Barbara, CA 93106-9530, USA}}

\newcommand{\UCD}{\affiliation{Department of Physics and Astronomy, University of California, Davis, 1 Shields Avenue, Davis, CA 95616-5270, USA}}

\newcommand{\OKC}{\affiliation{Oskar Klein Centre, Department of Astronomy, Stockholm University, Albanova University Centre, SE-106 91 Stockholm, Sweden}}

\newcommand{\Itagaki}{\affiliation{Itagaki Astronomical Observatory, Yamagata 990-2492, Japan}}

\newcommand{\CfA}{\affiliation{Center for Astrophysics \textbar{} Harvard \& Smithsonian, 60 Garden Street, Cambridge, MA 02138-1516, USA}}
\newcommand{\UA}{\affiliation{Steward Observatory, University of Arizona, 933 North Cherry Avenue, Tucson, AZ 85721-0065, USA}}

\newcommand{\UNC}{\affiliation{Department of Physics and Astronomy, University of North Carolina, 120 East Cameron Avenue, Chapel Hill, NC 27599, USA}}

\newcommand{\GeminiNorth}{\affiliation{Gemini Observatory, 670 North A`ohoku Place, Hilo, HI 96720-2700, USA}}
\newcommand{\Keck}{\affiliation{W.~M.~Keck Observatory, 65-1120 M\=amalahoa Highway, Kamuela, HI 96743-8431, USA}}
\newcommand{\UW}{\affiliation{Department of Astronomy, University of Washington, 3910 15th Avenue NE, Seattle, WA 98195-0002, USA}}

\newcommand{\USask}{\affiliation{Department of Physics \& Engineering Physics, University of Saskatchewan, 116 Science Place, Saskatoon, SK S7N 5E2, Canada}}

\newcommand{\Rutgers}{\affiliation{Department of Physics and Astronomy, Rutgers, the State University of New Jersey,\\136 Frelinghuysen Road, Piscataway, NJ 08854-8019, USA}}

\newcommand{\Macquarie}{\affiliation{School of Mathematical and Physical Sciences, Macquarie University, Sydney, NSW 2109, Australia}}
\newcommand{\AAARC}{\affiliation{Astrophysics and Space Technologies Research Centre, Macquarie University, Sydney, NSW 2109, Australia}}

\newcommand{\MSU}{\affiliation{Center for Data Intensive and Time Domain Astronomy, Department of Physics and Astronomy,\\Michigan State University, East Lansing, MI 48824, USA}}
\newcommand{\OSU}{\affiliation{Center for Cosmology and AstroParticle Physics, Ohio State University, 191 West Woodruff Avenue, Columbus, OH 43210-1117, USA}}
\newcommand{\IAIFI}{\affiliation{The NSF AI Institute for Artificial Intelligence and Fundamental Interactions, USA}}
\newcommand{\Warsaw}{\affiliation{Astronomical Observatory, University of Warsaw, Al.\ Ujazdowskie 4, 00-478 Warszawa, Poland}}
\newcommand{\SAAO}{\affiliation{South African Astronomical Observatory, P.O. Box 9, Observatory 7935, South Africa}}
\newcommand{\UFS}{\affiliation{Department of Physics, University of the Free State, P.O. Box 339, Bloemfontein 9300, South Africa}}
\newcommand{\UCT}{\affiliation{Department of Astronomy, University of Cape Town, Private Bag X3, Rondebosch 7701, South Africa}}
\newcommand{\Catalyst}{\altaffiliation{LSSTC Catalyst Fellow}}

%% file: main.bbl
\begin{thebibliography}{}\footnotesize
\expandafter\ifx\csname natexlab\endcsname\relax\def\natexlab#1{#1}\fi
\providecommand{\url}[1]{\href{#1}{#1}}
\providecommand{\dodoi}[1]{doi:~\href{http://doi.org/#1}{\nolinkurl{#1}}}

\bibitem[{Andrews {et~al.}(2023)Andrews, Sand, Valenti, Bostroem, Wyatt,
  Lundquist, Hoang, Jencson, Paraskeva, Janzen, Shrestha, Hosseinzadeh,
  Pearson, Dong, \& Meza}]{andrews_dlt40_2023}
Andrews, J.~E., Sand, D.~J., Valenti, S., {et~al.} 2023,
  \hypersetup{urlcolor=magenta}\href{https://www.wis-tns.org/object/2023bee/discovery-cert}{TNSTR},
  \hypersetup{urlcolor=blue}\href{https://ui.adsabs.harvard.edu/abs/2023TNSTR.256....1A}{256,
  1}

\bibitem[{Ashall {et~al.}(2022)Ashall, Lu, Shappee, Burns, Hsiao, Kumar,
  Morrell, Phillips, Shahbandeh, Baron, Boutsia, Brown, DerKacy, Galbany,
  Hoeflich, Krisciunas, Mazzali, Piro, Stritzinger, \&
  Suntzeff}]{ashall_speed_2022}
Ashall, C., Lu, J., Shappee, B.~J., {et~al.} 2022,
  \hypersetup{urlcolor=magenta}\href{https://dx.doi.org/10.3847/2041-8213/ac7235}{ApJL},
  \hypersetup{urlcolor=blue}\href{https://ui.adsabs.harvard.edu/abs/2022ApJL..932L...2A}{932,
  L2}

\bibitem[{{Astropy Collaboration} {et~al.}(2022){Astropy Collaboration},
  {Price-Whelan}, Lim, Earl, Starkman, Bradley, Shupe, Patil, Corrales,
  Brasseur, N{\"o}the, Donath, Tollerud, Morris, Ginsburg, Vaher, Weaver,
  Tocknell, Jamieson, {van Kerkwijk}, Robitaille, Merry, Bachetti, G{\"u}nther,
  Aldcroft, {Alvarado-Montes}, Archibald, B{\'o}di, Bapat, Barentsen,
  Baz{\'a}n, Biswas, Boquien, Burke, Cara, Cara, Conroy, Conseil, Craig, Cross,
  Cruz, D'Eugenio, Dencheva, Devillepoix, Dietrich, Eigenbrot, Erben, Ferreira,
  {Foreman-Mackey}, Fox, Freij, Garg, Geda, Glattly, Gondhalekar, Gordon,
  Grant, Greenfield, Groener, Guest, Gurovich, Handberg, Hart,
  {Hatfield-Dodds}, Homeier, Hosseinzadeh, Jenness, Jones, Joseph, Kalmbach,
  Karamehmetoglu, Ka{\l}uszy{\'n}ski, Kelley, Kern, Kerzendorf, Koch, Kulumani,
  Lee, Ly, Ma, MacBride, Maljaars, Muna, Murphy, Norman, O'Steen, Oman,
  Pacifici, Pascual, {Pascual-Granado}, Patil, Perren, Pickering, Rastogi,
  Roulston, Ryan, Rykoff, Sabater, Sakurikar, Salgado, Sanghi, Saunders,
  Savchenko, Schwardt, {Seifert-Eckert}, Shih, Jain, Shukla, Sick, Simpson,
  Singanamalla, Singer, Singhal, Sinha, Sip{\H o}cz, Spitler, Stansby,
  Streicher, {\v S}umak, Swinbank, Taranu, Tewary, Tremblay, de~{Val-Borro},
  Van~Kooten, Vasovi{\'c}, Verma, {de Miranda Cardoso}, Williams, Wilson,
  Winkel, {Wood-Vasey}, Xue, Yoachim, Zhang, Zonca, \& {Astropy Project
  Contributors}}]{astropycollaboration_astropy_2022}
{Astropy Collaboration}, {Price-Whelan}, A.~M., Lim, P.~L., {et~al.} 2022,
  \hypersetup{urlcolor=magenta}\href{https://dx.doi.org/10.3847/1538-4357/ac7c74}{ApJ},
  \hypersetup{urlcolor=blue}\href{https://ui.adsabs.harvard.edu/abs/2022ApJ...935..167A}{935,
  167}

\bibitem[{Barbary(2016)}]{barbary_extinction_2016}
Barbary, K. 2016, Extinction v0.3.0, Zenodo,
  \hypersetup{urlcolor=magenta}doi:\href{https://doi.org/10.5281/zenodo.804967}{10.5281/zenodo.804967}

\bibitem[{Barbary {et~al.}(2022)Barbary, Bailey, Barentsen, Barclay, Biswas,
  Boone, Craig, Feindt, Friesen, Goldstein, Jha, Jones, Mondon,
  Papadogiannakis, Perrefort, Pierel, Rodney, Rose, Saunders, Sip{\H o}cz,
  Sofiatti, Thomas, {van Santen}, Vincenzi, Wang, \&
  {Wood-Vasey}}]{barbary_sncosmo_2022}
Barbary, K., Bailey, S., Barentsen, G., {et~al.} 2022, {{SNCosmo}} v2.8.0,
  Zenodo,
  \hypersetup{urlcolor=magenta}doi:\href{https://doi.org/10.5281/zenodo.6363879}{10.5281/zenodo.6363879}

\bibitem[{Bostroem {et~al.}(2021)Bostroem, Jha, Randriamampandry, Valenti,
  Sand, Wyatt, Lundquist, Andrews, Jencson, Dong, Janzen, Pearson,
  Hosseinzadeh, \& Meza}]{bostroem_global_2021}
Bostroem, K.~A., Jha, S.~W., Randriamampandry, S., {et~al.} 2021,
  \hypersetup{urlcolor=magenta}\href{https://www.wis-tns.org/object/2021aefx/classification-cert}{TNSCR},
  \hypersetup{urlcolor=blue}\href{https://ui.adsabs.harvard.edu/abs/2021TNSCR3888....1B}{3888,
  1}

\bibitem[{Boty{\'a}nszki {et~al.}(2018)Boty{\'a}nszki, Kasen, \&
  Plewa}]{botyanszki_multidimensional_2018}
Boty{\'a}nszki, J., Kasen, D., \& Plewa, T. 2018,
  \hypersetup{urlcolor=magenta}\href{https://dx.doi.org/10.3847/2041-8213/aaa07b}{ApJL},
  \hypersetup{urlcolor=blue}\href{https://ui.adsabs.harvard.edu/abs/2018ApJL..852L...6B}{852,
  L6}

\bibitem[{Bradley {et~al.}(2022)Bradley, Sip{\H o}cz, Robitaille, Tollerud,
  Vin{\'i}cius, Deil, Barbary, Wilson, Busko, Donath, G{\"u}nther, Cara, Lim,
  Me{\ss}linger, Conseil, Bostroem, Droettboom, Bray, Bratholm, Barentsen,
  Craig, Rathi, Pascual, Perren, Georgiev, {de Val-Borro}, Kerzendorf, Bach,
  Quint, \& Souchereau}]{bradley_astropy_2022}
Bradley, L., Sip{\H o}cz, B., Robitaille, T., {et~al.} 2022,
  Astropy/{{Photutils}} v1.5.0, Zenodo,
  \hypersetup{urlcolor=magenta}doi:\href{https://doi.org/10.5281/zenodo.6825092}{10.5281/zenodo.6825092}

\bibitem[{Breeveld {et~al.}(2010)Breeveld, Curran, Hoversten, Koch, Landsman,
  Marshall, Page, Poole, Roming, Smith, Still, Yershov, Blustin, Brown,
  Gronwall, Holland, Kuin, McGowan, Rosen, Boyd, Broos, Carter, Chester,
  Hancock, Huckle, Immler, Ivanushkina, Kennedy, Mason, Morgan, Oates,
  De~Pasquale, Schady, Siegel, \& Vanden~Berk}]{breeveld_further_2010}
Breeveld, A.~A., Curran, P.~A., Hoversten, E.~A., {et~al.} 2010,
  \hypersetup{urlcolor=magenta}\href{https://dx.doi.org/10.1111/j.1365-2966.2010.16832.x}{MNRAS},
  \hypersetup{urlcolor=blue}\href{https://ui.adsabs.harvard.edu/abs/2010MNRAS.406.1687B}{406,
  1687}

\bibitem[{Brown {et~al.}(2012)Brown, Dawson, Harris, Olmstead, Milne, \&
  Roming}]{brown_constraints_2012}
Brown, P.~J., Dawson, K.~S., Harris, D.~W., {et~al.} 2012,
  \hypersetup{urlcolor=magenta}\href{https://dx.doi.org/10.1088/0004-637X/749/1/18}{ApJ},
  \hypersetup{urlcolor=blue}\href{https://ui.adsabs.harvard.edu/abs/2012ApJ...749...18B}{749,
  18}

\bibitem[{Brown {et~al.}(2013)Brown, Baliber, Bianco, Bowman, Burleson, Conway,
  Crellin, Depagne, De~Vera, Dilday, Dragomir, Dubberley, Eastman, Elphick,
  Falarski, Foale, Ford, Fulton, Garza, Gomez, Graham, Greene, Haldeman,
  Hawkins, Haworth, Haynes, Hidas, Hjelstrom, Howell, Hygelund, Lister,
  Lobdill, Martinez, Mullins, Norbury, Parrent, Paulson, Petry, Pickles,
  Posner, Rosing, Ross, Sand, Saunders, Shobbrook, Shporer, Street, Thomas,
  Tsapras, Tufts, Valenti, Vander~Horst, Walker, White, \&
  Willis}]{brown_cumbres_2013}
Brown, T.~M., Baliber, N., Bianco, F.~B., {et~al.} 2013,
  \hypersetup{urlcolor=magenta}\href{https://dx.doi.org/10.1086/673168}{PASP},
  \hypersetup{urlcolor=blue}\href{https://ui.adsabs.harvard.edu/abs/2013PASP..125.1031B}{125,
  1031}

\bibitem[{Buckley {et~al.}(2006)Buckley, Burgh, Cottrell, Nordsieck,
  O'Donoghue, \& Williams}]{buckley_status_2006}
Buckley, D. A.~H., Burgh, E.~B., Cottrell, P.~L., {et~al.} 2006,
  \hypersetup{urlcolor=magenta}\href{https://dx.doi.org/10.1117/12.673838}{Proc.\
  SPIE},
  \hypersetup{urlcolor=blue}\href{https://ui.adsabs.harvard.edu/abs/2006SPIE.6269E..0AB}{6269,
  0A}

\bibitem[{Bufano {et~al.}(2014)Bufano, Pignata, Bersten, Mazzali, Ryder,
  Margutti, Milisavljevic, Morelli, Benetti, Cappellaro, {Gonzalez-Gaitan},
  {Romero-Ca{\~n}izales}, Stritzinger, Walker, Anderson, Contreras, {de
  Jaeger}, F{\"o}rster, Gutierrez, Hamuy, Hsiao, Morrell, Olivares~E., Paillas,
  Parker, Pian, Pickering, Sanders, Stockdale, Turatto, Valenti, Fesen, Maza,
  Nomoto, Phillips, \& Soderberg}]{bufano_sn_2014}
Bufano, F., Pignata, G., Bersten, M., {et~al.} 2014,
  \hypersetup{urlcolor=magenta}\href{https://dx.doi.org/10.1093/mnras/stu065}{MNRAS},
  \hypersetup{urlcolor=blue}\href{https://ui.adsabs.harvard.edu/abs/2014MNRAS.439.1807B}{439,
  1807}

\bibitem[{Burke {et~al.}(2022{\natexlab{\hspace{0pt}a}})Burke, Howell, Sand, \&
  Hosseinzadeh}]{burke_companion_2022}
Burke, J., Howell, D.~A., Sand, D.~J., \& Hosseinzadeh, G.
  2022{\natexlab{\hspace{0pt}a}},
  \hypersetup{urlcolor=magenta}\href{https://arxiv.org/abs/2208.11201}{arXiv}{:}\hypersetup{urlcolor=blue}\href{https://ui.adsabs.harvard.edu/abs/2022arXiv220811201B}{2208.11201}

\bibitem[{Burke {et~al.}(2021)Burke, Howell, Sarbadhicary, Sand, Amaro,
  Hiramatsu, McCully, Pellegrino, Andrews, Brown, Itagaki, Shahbandeh,
  Bostroem, Chomiuk, Hsiao, Smith, \& Valenti}]{burke_bright_2021}
Burke, J., Howell, D.~A., Sarbadhicary, S.~K., {et~al.} 2021,
  \hypersetup{urlcolor=magenta}\href{https://dx.doi.org/10.3847/1538-4357/ac126b}{ApJ},
  \hypersetup{urlcolor=blue}\href{https://ui.adsabs.harvard.edu/abs/2021ApJ...919..142B}{919,
  142}

\bibitem[{Burke {et~al.}(2022{\natexlab{\hspace{0pt}b}})Burke, Howell, Sand,
  Amaro, Brown, Andrews, Bostroem, Dong, Haislip, Hiramatsu, Hosseinzadeh,
  Kouprianov, Lundquist, McCully, Pellegrino, Reichart, Tartaglia, Valenti, \&
  Yang}]{burke_early_2022}
Burke, J., Howell, D.~A., Sand, D.~J., {et~al.} 2022{\natexlab{\hspace{0pt}b}},
  \hypersetup{urlcolor=magenta}\href{https://arxiv.org/abs/2207.07681}{arXiv}{:}\hypersetup{urlcolor=blue}\href{https://ui.adsabs.harvard.edu/abs/2022arXiv220707681B}{2207.07681}

\bibitem[{Cao {et~al.}(2015)Cao, Kulkarni, Howell, {Gal-Yam}, Kasliwal,
  Valenti, Johansson, Amanullah, Goobar, Sollerman, Taddia, Horesh, Sagiv,
  Cenko, Nugent, Arcavi, Surace, Wo{\'z}niak, Moody, Rebbapragada, Bue, \&
  Gehrels}]{cao_strong_2015}
Cao, Y., Kulkarni, S.~R., Howell, D.~A., {et~al.} 2015,
  \hypersetup{urlcolor=magenta}\href{https://dx.doi.org/10.1038/nature14440}{Natur},
  \hypersetup{urlcolor=blue}\href{https://ui.adsabs.harvard.edu/abs/2015Natur.521..328C}{521,
  328}

\bibitem[{Chambers {et~al.}(2016)Chambers, Magnier, Metcalfe, Flewelling,
  Huber, Waters, Denneau, Draper, Farrow, Finkbeiner, Holmberg, Koppenhoefer,
  Price, Rest, Saglia, Schlafly, Smartt, Sweeney, Wainscoat, Burgett, Chastel,
  Grav, Heasley, Hodapp, Jedicke, Kaiser, Kudritzki, Luppino, Lupton, Monet,
  Morgan, Onaka, Shiao, Stubbs, Tonry, White, Ba{\~n}ados, Bell, Bender,
  Bernard, Boegner, Boffi, Botticella, Calamida, Casertano, Chen, Chen, Cole,
  Deacon, Frenk, Fitzsimmons, Gezari, Gibbs, Goessl, Goggia, Gourgue, Goldman,
  Grant, Grebel, Hambly, Hasinger, Heavens, Heckman, Henderson, Henning,
  Holman, Hopp, Ip, Isani, Jackson, Keyes, Koekemoer, Kotak, Le, Liska, Long,
  Lucey, Liu, Martin, Masci, McLean, Mindel, Misra, Morganson, Murphy, Obaika,
  Narayan, {Nieto-Santisteban}, Norberg, Peacock, Pier, Postman, Primak, Rae,
  Rai, Riess, Riffeser, Rix, R{\"o}ser, Russel, Rutz, Schilbach, Schultz,
  Scolnic, Strolger, Szalay, Seitz, Small, Smith, Soderblom, Taylor, Thomson,
  Taylor, Thakar, Thiel, Thilker, Unger, Urata, Valenti, Wagner, Walder,
  Walter, Watters, Werner, {Wood-Vasey}, \& Wyse}]{chambers_pan-starrs1_2016}
Chambers, K.~C., Magnier, E.~A., Metcalfe, N., {et~al.} 2016,
  \hypersetup{urlcolor=magenta}\href{https://arxiv.org/abs/1612.05560}{arXiv}{:}\hypersetup{urlcolor=blue}\href{https://ui.adsabs.harvard.edu/abs/2016arXiv161205560C}{1612.05560}

\bibitem[{Chevalier(1982)}]{chevalier_radio_1982}
Chevalier, R.~A. 1982,
  \hypersetup{urlcolor=magenta}\href{https://dx.doi.org/10.1086/160167}{ApJ},
  \hypersetup{urlcolor=blue}\href{https://ui.adsabs.harvard.edu/abs/1982ApJ...259..302C}{259,
  302}

\bibitem[{Chomiuk {et~al.}(2012)Chomiuk, Soderberg, Moe, Chevalier, Rupen,
  Badenes, Margutti, Fransson, Fong, \& Dittmann}]{chomiuk_evla_2012}
Chomiuk, L., Soderberg, A.~M., Moe, M., {et~al.} 2012,
  \hypersetup{urlcolor=magenta}\href{https://dx.doi.org/10.1088/0004-637X/750/2/164}{ApJ},
  \hypersetup{urlcolor=blue}\href{https://ui.adsabs.harvard.edu/abs/2012ApJ...750..164C}{750,
  164}

\bibitem[{Chomiuk {et~al.}(2016)Chomiuk, Soderberg, Chevalier, Bruzewski,
  Foley, Parrent, Strader, Badenes, Fransson, Kamble, Margutti, Rupen, \&
  Simon}]{chomiuk_deep_2016}
Chomiuk, L., Soderberg, A.~M., Chevalier, R.~A., {et~al.} 2016,
  \hypersetup{urlcolor=magenta}\href{https://dx.doi.org/10.3847/0004-637X/821/2/119}{ApJ},
  \hypersetup{urlcolor=blue}\href{https://ui.adsabs.harvard.edu/abs/2016ApJ...821..119C}{821,
  119}

\bibitem[{Clemens {et~al.}(2004)Clemens, Crain, \&
  Anderson}]{clemens_goodman_2004}
Clemens, J.~C., Crain, J.~A., \& Anderson, R. 2004,
  \hypersetup{urlcolor=magenta}\href{https://dx.doi.org/10.1117/12.550069}{Proc.\
  SPIE},
  \hypersetup{urlcolor=blue}\href{https://ui.adsabs.harvard.edu/abs/2004SPIE.5492..331C}{5492,
  331}

\bibitem[{Conley {et~al.}(2008)Conley, Sullivan, Hsiao, Guy, Astier, Balam,
  Balland, Basa, Carlberg, {D. Fouchez}, Hardin, Howell, Hook, Pain, Perrett,
  Pritchet, \& Regnault}]{conley_sifto:_2008}
Conley, A., Sullivan, M., Hsiao, E.~Y., {et~al.} 2008,
  \hypersetup{urlcolor=magenta}\href{https://dx.doi.org/10.1086/588518}{ApJ},
  \hypersetup{urlcolor=blue}\href{https://ui.adsabs.harvard.edu/abs/2008ApJ...681..482C}{681,
  482}

\bibitem[{Crawford {et~al.}(2010)Crawford, Still, Schellart, Balona, Buckley,
  Dugmore, Gulbis, Kniazev, Kotze, Loaring, Nordsieck, Pickering, Potter,
  Romero~Colmenero, Vaisanen, Williams, \& Zietsman}]{crawford_pysalt:_2010}
Crawford, S.~M., Still, M., Schellart, P., {et~al.} 2010,
  \hypersetup{urlcolor=magenta}\href{https://dx.doi.org/10.1117/12.857000}{Proc.\
  SPIE},
  \hypersetup{urlcolor=blue}\href{https://ui.adsabs.harvard.edu/abs/2010SPIE.7737E..25C}{7737,
  25}

\bibitem[{Deckers {et~al.}(2022)Deckers, Maguire, Magee, Dimitriadis, Smith,
  {Sainz~de~Murieta}, Miller, Goobar, Nordin, Rigault, Bellm, Coughlin, Laher,
  Shupe, Graham, Kasliwal, \& Walters}]{deckers_constraining_2022}
Deckers, M., Maguire, K., Magee, M.~R., {et~al.} 2022,
  \hypersetup{urlcolor=magenta}\href{https://dx.doi.org/10.1093/mnras/stac558}{MNRAS},
  \hypersetup{urlcolor=blue}\href{https://ui.adsabs.harvard.edu/abs/2022MNRAS.512.1317D}{512,
  1317}

\bibitem[{Dessart {et~al.}(2014)Dessart, Blondin, Hillier, \&
  Khokhlov}]{dessart_constraints_2014}
Dessart, L., Blondin, S., Hillier, D.~J., \& Khokhlov, A. 2014,
  \hypersetup{urlcolor=magenta}\href{https://dx.doi.org/10.1093/mnras/stu598}{MNRAS},
  \hypersetup{urlcolor=blue}\href{https://ui.adsabs.harvard.edu/abs/2014MNRAS.441..532D}{441,
  532}

\bibitem[{Dessart {et~al.}(2020)Dessart, Leonard, \&
  Prieto}]{dessart_spectral_2020}
Dessart, L., Leonard, D.~C., \& Prieto, J.~L. 2020,
  \hypersetup{urlcolor=magenta}\href{https://dx.doi.org/10.1051/0004-6361/202037854}{A\&A},
  \hypersetup{urlcolor=blue}\href{https://ui.adsabs.harvard.edu/abs/2020A&A...638A..80D}{638,
  A80}

\bibitem[{Dimitriadis {et~al.}(2019{\natexlab{\hspace{0pt}a}})Dimitriadis,
  Foley, Rest, Kasen, Piro, Polin, Jones, Villar, Narayan, Coulter, Kilpatrick,
  Pan, {Rojas-Bravo}, Fox, Jha, Nugent, Riess, Scolnic, Drout, Barentsen,
  Dotson, {Gully-Santiago}, Hedges, Cody, Barclay, Howell, Garnavich, Tucker,
  Shaya, Mushotzky, Olling, Margheim, Zenteno, Coughlin, Cleve, Cardoso,
  Larson, {McCalmont-Everton}, Peterson, Ross, Reedy, Osborne, McGinn, Kohnert,
  Migliorini, Wheaton, Spencer, Labonde, Castillo, Beerman, Steward, Hanley,
  Larsen, Gangopadhyay, Kloetzel, Weschler, Nystrom, Moffatt, Redick, Griest,
  Packard, Muszynski, Kampmeier, Bjella, Flynn, Elsaesser, Chambers,
  Flewelling, Huber, Magnier, Waters, Schultz, Bulger, Lowe, Willman, Smartt,
  Smith, Points, Strampelli, Brimacombe, Chen, Mu{\~n}oz, Mutel, Shields,
  Vallely, Villanueva, Li, Wang, Zhang, Lin, Mo, Zhao, Sai, Zhang, Zhang,
  Zhang, Wang, Zhang, Baron, DerKacy, Li, Chen, Xiang, Rui, Wang, Huang, Li,
  Hosseinzadeh, Howell, Arcavi, Hiramatsu, Burke, Valenti, Tonry, Denneau,
  Heinze, Weiland, Stalder, Vink{\'o}, S{\'a}rneczky, P{\'a}l, B{\'o}di,
  Bogn{\'a}r, Cs{\'a}k, Cseh, Cs{\"o}rnyei, Hanyecz, Ign{\'a}cz, Kalup,
  {K{\"o}nyves-T{\'o}th}, Kriskovics, Ordasi, Rajmon, S{\'o}dor, Szab{\'o},
  Szak{\'a}ts, Zsidi, Williams, Nordin, Cartier, Frohmaier, Galbany,
  Guti{\'e}rrez, Hook, Inserra, Smith, Sand, Andrews, Smith, Bilinski, {and},
  {and}, {and}, {and}, {and}, \& {and}}]{dimitriadis_k2_2019}
Dimitriadis, G., Foley, R.~J., Rest, A., {et~al.}
  2019{\natexlab{\hspace{0pt}a}},
  \hypersetup{urlcolor=magenta}\href{https://dx.doi.org/10.3847/2041-8213/aaedb0}{ApJL},
  \hypersetup{urlcolor=blue}\href{https://ui.adsabs.harvard.edu/abs/2019ApJL..870L...1D}{870,
  L1}

\bibitem[{Dimitriadis {et~al.}(2019{\natexlab{\hspace{0pt}b}})Dimitriadis,
  {Rojas-Bravo}, Kilpatrick, Foley, Piro, Brown, Guhathakurta, Quirk, Rest,
  Strampelli, Tucker, \& Villar}]{dimitriadis_nebular_2019}
Dimitriadis, G., {Rojas-Bravo}, C., Kilpatrick, C.~D., {et~al.}
  2019{\natexlab{\hspace{0pt}b}},
  \hypersetup{urlcolor=magenta}\href{https://dx.doi.org/10.3847/2041-8213/aaf9b1}{ApJL},
  \hypersetup{urlcolor=blue}\href{https://ui.adsabs.harvard.edu/abs/2019ApJ...870L..14D}{870,
  L14}

\bibitem[{Dimitriadis {et~al.}(2023)Dimitriadis, Maguire, Karambelkar, Lebron,
  Liu (刘~畅), Kozyreva, Miller, {Ridden-Harper}, Anderson, Chen, Coughlin,
  Valle, Drake, Galbany, Gromadzki, Groom, Guti{\'e}rrez, Ihanec, Inserra,
  Johansson, {M{\"u}ller-Bravo}, Nicholl, Polin, Rusholme, Schulze, Sollerman,
  Srivastav, Taggart, Wang, Yang (杨~轶), \& Young}]{dimitriadis_sn_2023}
Dimitriadis, G., Maguire, K., Karambelkar, V.~R., {et~al.} 2023,
  \hypersetup{urlcolor=magenta}\href{https://dx.doi.org/10.1093/mnras/stad536}{MNRAS},
  \hypersetup{urlcolor=blue}\href{https://ui.adsabs.harvard.edu/abs/2023MNRAS.521.1162D}{521,
  1162}

\bibitem[{Fabricant {et~al.}(2019)Fabricant, Fata, Epps, Gauron, Mueller,
  Zajac, Amato, Barberis, Bergner, Brennan, Brown, Chilingarian, Geary,
  Kradinov, McLeod, Smith, \& Woods}]{fabricant_binospec:_2019}
Fabricant, D., Fata, R., Epps, H., {et~al.} 2019,
  \hypersetup{urlcolor=magenta}\href{https://dx.doi.org/10.1088/1538-3873/ab1d78}{PASP},
  \hypersetup{urlcolor=blue}\href{https://ui.adsabs.harvard.edu/abs/2019PASP..131075004F}{131,
  075004}

\bibitem[{Falco {et~al.}(1999)Falco, Kurtz, Geller, Huchra, Peters, Berlind,
  Mink, Tokarz, \& Elwell}]{falco_updated_1999}
Falco, E.~E., Kurtz, M.~J., Geller, M.~J., {et~al.} 1999,
  \hypersetup{urlcolor=magenta}\href{https://dx.doi.org/10.1086/316343}{PASP},
  \hypersetup{urlcolor=blue}\href{https://ui.adsabs.harvard.edu/abs/1999PASP..111..438F}{111,
  438}

\bibitem[{Fitzpatrick(1999)}]{fitzpatrick_correcting_1999}
Fitzpatrick, E.~L. 1999,
  \hypersetup{urlcolor=magenta}\href{https://dx.doi.org/10.1086/316293}{PASP},
  \hypersetup{urlcolor=blue}\href{https://ui.adsabs.harvard.edu/abs/1999PASP..111...63F}{111,
  63}

\bibitem[{Folatelli {et~al.}(2012)Folatelli, Phillips, Morrell, Tanaka, Maeda,
  {Ken'ichi Nomoto}, Stritzinger, Burns, Hamuy, Mazzali, Boldt, {Abdo
  Campillay}, Contreras, Gonz{\'a}lez, Roth, Salgado, Freedman, Madore,
  Persson, \& Suntzeff}]{folatelli_unburned_2012}
Folatelli, G., Phillips, M.~M., Morrell, N., {et~al.} 2012,
  \hypersetup{urlcolor=magenta}\href{https://dx.doi.org/10.1088/0004-637X/745/1/74}{ApJ},
  \hypersetup{urlcolor=blue}\href{https://ui.adsabs.harvard.edu/abs/2012ApJ...745...74F}{745,
  74}

\bibitem[{{Foreman-Mackey}(2016)}]{foreman-mackey_corner.py:_2016}
{Foreman-Mackey}, D. 2016,
  \hypersetup{urlcolor=magenta}\href{https://dx.doi.org/10.21105/joss.00024}{JOSS},
  \hypersetup{urlcolor=blue}\href{https://ui.adsabs.harvard.edu/abs/2016JOSS....1...24F}{1,
  24}

\bibitem[{{Foreman-Mackey} {et~al.}(2013){Foreman-Mackey}, Hogg, Lang, \&
  Goodman}]{foreman-mackey_emcee_2013}
{Foreman-Mackey}, D., Hogg, D.~W., Lang, D., \& Goodman, J. 2013,
  \hypersetup{urlcolor=magenta}\href{https://dx.doi.org/10.1086/670067}{PASP},
  \hypersetup{urlcolor=blue}\href{https://ui.adsabs.harvard.edu/abs/2013PASP..125..306F}{125,
  306}

\bibitem[{Fransson {et~al.}(1996)Fransson, Lundqvist, \&
  Chevalier}]{fransson_circumstellar_1996}
Fransson, C., Lundqvist, P., \& Chevalier, R.~A. 1996,
  \hypersetup{urlcolor=magenta}\href{https://dx.doi.org/10.1086/177119}{ApJ},
  \hypersetup{urlcolor=blue}\href{https://ui.adsabs.harvard.edu/abs/1996ApJ...461..993F}{461,
  993}

\bibitem[{Graham {et~al.}(2017)Graham, Kumar, Hosseinzadeh, Hiramatsu, Arcavi,
  Howell, Valenti, Sand, Parrent, McCully, \&
  Filippenko}]{graham_nebular-phase_2017}
Graham, M.~L., Kumar, S., Hosseinzadeh, G., {et~al.} 2017,
  \hypersetup{urlcolor=magenta}\href{https://dx.doi.org/10.1093/mnras/stx2224}{MNRAS},
  \hypersetup{urlcolor=blue}\href{https://ui.adsabs.harvard.edu/abs/2017MNRAS.472.3437G}{472,
  3437}

\bibitem[{Green {et~al.}(1995)Green, Schmidt, Oey, Wittman, \&
  Hall}]{green_steward_1995}
Green, R., Schmidt, G., Oey, S., Wittman, D., \& Hall, P. 1995, Steward
  {{Observatory}} 2.3-m {{Boller}} and {{Chivens Spectrograph Manual}},
  \hypersetup{urlcolor=magenta}\url{http://james.as.arizona.edu/~psmith/90inch/bcman/html/bcman.html}

\bibitem[{Henden {et~al.}(2009)Henden, Welch, Terrell, \&
  Levine}]{henden_aavso_2009}
Henden, A.~A., Welch, D.~L., Terrell, D., \& Levine, S.~E. 2009, AAS,
  \hypersetup{urlcolor=blue}\href{http://adsabs.harvard.edu/abs/2009AAS...21440702H}{41,
  407.02}

\bibitem[{Hoeflich(2017)}]{hoeflich_explosion_2017}
Hoeflich, P. 2017, in Handbook of {{Supernovae}}, ed. A.~W. Alsabti \&
  P.~Murdin ({Cham}:
  \hypersetup{urlcolor=magenta}\href{https://doi.org/10.1007/978-3-319-20794-0_56-1}{{Springer}})

\bibitem[{Horesh {et~al.}(2012)Horesh, Kulkarni, Fox, Carpenter, Kasliwal,
  Ofek, Quimby, {Gal-Yam}, Cenko, de~Bruyn, Kamble, Wijers, van~der Horst,
  Kouveliotou, Podsiadlowski, Sullivan, Maguire, Howell, Nugent, Gehrels, Law,
  Poznanski, \& Shara}]{horesh_early_2012}
Horesh, A., Kulkarni, S.~R., Fox, D.~B., {et~al.} 2012,
  \hypersetup{urlcolor=magenta}\href{https://dx.doi.org/10.1088/0004-637X/746/1/21}{ApJ},
  \hypersetup{urlcolor=blue}\href{https://ui.adsabs.harvard.edu/abs/2012ApJ...746...21H}{746,
  21}

\bibitem[{Hosseinzadeh {et~al.}(2023{\natexlab{\hspace{0pt}a}})Hosseinzadeh,
  Bostroem, \& Gomez}]{hosseinzadeh_light_2023}
Hosseinzadeh, G., Bostroem, K.~A., \& Gomez, S. 2023{\natexlab{\hspace{0pt}a}},
  Light {{Curve Fitting}} v0.8.0, Zenodo,
  \hypersetup{urlcolor=magenta}doi:\href{https://doi.org/10.5281/zenodo.7872772}{10.5281/zenodo.7872772}

\bibitem[{Hosseinzadeh {et~al.}(2017)Hosseinzadeh, Sand, Valenti, Brown,
  Howell, McCully, Kasen, Arcavi, Bostroem, Tartaglia, Hsiao, Davis,
  Shahbandeh, \& Stritzinger}]{hosseinzadeh_early_2017}
Hosseinzadeh, G., Sand, D.~J., Valenti, S., {et~al.} 2017,
  \hypersetup{urlcolor=magenta}\href{https://dx.doi.org/10.3847/2041-8213/aa8402}{ApJL},
  \hypersetup{urlcolor=blue}\href{https://ui.adsabs.harvard.edu/abs/2017ApJL..845L..11H}{845,
  L11}

\bibitem[{Hosseinzadeh {et~al.}(2022)Hosseinzadeh, Sand, Lundqvist, Andrews,
  Bostroem, Dong, Janzen, Jencson, Lundquist, Retamal, Pearson, Valenti, Wyatt,
  Burke, Howell, McCully, Newsome, Gonzalez, Pellegrino, Terreran, Kwok, Jha,
  Strader, Kundu, Ryder, Haislip, Kouprianov, \&
  Reichart}]{hosseinzadeh_constraining_2022}
Hosseinzadeh, G., Sand, D.~J., Lundqvist, P., {et~al.} 2022,
  \hypersetup{urlcolor=magenta}\href{https://dx.doi.org/10.3847/2041-8213/ac7cef}{ApJL},
  \hypersetup{urlcolor=blue}\href{https://ui.adsabs.harvard.edu/abs/2022ApJL..933L..45H}{933,
  L45}

\bibitem[{Hosseinzadeh {et~al.}(2023{\natexlab{\hspace{0pt}b}})Hosseinzadeh,
  Newsome, Sand, Farah, Howell, McCully, Gonzalez, Pellegrino, \&
  Terreran}]{hosseinzadeh_global_2023}
Hosseinzadeh, G., Newsome, M., Sand, D.~J., {et~al.}
  2023{\natexlab{\hspace{0pt}b}},
  \hypersetup{urlcolor=magenta}\href{https://ui.adsabs.harvard.edu/abs/2023TNSCR.277....1H}{TNSCR},
  \hypersetup{urlcolor=blue}\href{https://ui.adsabs.harvard.edu/abs/2023TNSCR.277....1H}{277,
  1}

\bibitem[{Hunter(2007)}]{hunter_matplotlib:_2007}
Hunter, J.~D. 2007,
  \hypersetup{urlcolor=magenta}\href{https://dx.doi.org/10.1109/MCSE.2007.55}{CSE},
  \hypersetup{urlcolor=blue}\href{https://ui.adsabs.harvard.edu/abs/2007CSE.....9...90H}{9,
  90}

\bibitem[{Iben \& Tutukov(1984)}]{iben_supernovae_1984}
Iben, I., \& Tutukov, A.~V. 1984,
  \hypersetup{urlcolor=magenta}\href{https://dx.doi.org/10.1086/190932}{ApJS},
  \hypersetup{urlcolor=blue}\href{https://ui.adsabs.harvard.edu/abs/1984ApJS...54..335I}{54,
  335}

\bibitem[{Jha {et~al.}(2019)Jha, Maguire, \& Sullivan}]{jha_observational_2019}
Jha, S.~W., Maguire, K., \& Sullivan, M. 2019,
  \hypersetup{urlcolor=magenta}\href{https://dx.doi.org/10.1038/s41550-019-0858-0}{NatAs},
  \hypersetup{urlcolor=blue}\href{https://ui.adsabs.harvard.edu/abs/2019NatAs...3..706J}{3,
  706}

\bibitem[{Jiang {et~al.}(2017)Jiang, Doi, Maeda, Shigeyama, Nomoto, Yasuda,
  Jha, Tanaka, Morokuma, Tominaga, Ivezi{\'c}, {Ruiz-Lapuente}, Stritzinger,
  Mazzali, Ashall, Mould, Baade, Suzuki, Connolly, Patat, Wang, Yoachim, Jones,
  Furusawa, \& Miyazaki}]{jiang_hybrid_2017}
Jiang, J., Doi, M., Maeda, K., {et~al.} 2017,
  \hypersetup{urlcolor=magenta}\href{https://dx.doi.org/10.1038/nature23908}{Natur},
  \hypersetup{urlcolor=blue}\href{https://ui.adsabs.harvard.edu/abs/2017Natur.550...80J}{550,
  80}

\bibitem[{Jordi {et~al.}(2006)Jordi, Grebel, \& Ammon}]{jordi_empirical_2006}
Jordi, K., Grebel, E.~K., \& Ammon, K. 2006,
  \hypersetup{urlcolor=magenta}\href{https://dx.doi.org/10.1051/0004-6361:20066082}{A\&A},
  \hypersetup{urlcolor=blue}\href{https://ui.adsabs.harvard.edu/abs/2006A&A...460..339J}{460,
  339}

\bibitem[{Kansky {et~al.}(2019)Kansky, Chilingarian, Fabricant, Matthews,
  Moran, Paegert, Duane~Gibson, Porter, \& Roll}]{kansky_binospec_2019}
Kansky, J., Chilingarian, I., Fabricant, D., {et~al.} 2019,
  \hypersetup{urlcolor=magenta}\href{https://dx.doi.org/10.1088/1538-3873/ab1ceb}{PASP},
  \hypersetup{urlcolor=blue}\href{https://ui.adsabs.harvard.edu/abs/2019PASP..131075005K}{131,
  075005}

\bibitem[{Kasen(2010)}]{kasen_seeing_2010}
Kasen, D. 2010,
  \hypersetup{urlcolor=magenta}\href{https://dx.doi.org/10.1088/0004-637X/708/2/1025}{ApJ},
  \hypersetup{urlcolor=blue}\href{https://ui.adsabs.harvard.edu/abs/2010ApJ...708.1025K}{708,
  1025}

\bibitem[{Kenworthy {et~al.}(2021)Kenworthy, Jones, Dai, Kessler, Scolnic,
  Brout, Siebert, Pierel, Dettman, Dimitriadis, Foley, Jha, Pan, Riess, Rodney,
  \& {Rojas-Bravo}}]{kenworthy_salt3_2021}
Kenworthy, W.~D., Jones, D.~O., Dai, M., {et~al.} 2021,
  \hypersetup{urlcolor=magenta}\href{https://dx.doi.org/10.3847/1538-4357/ac30d8}{ApJ},
  \hypersetup{urlcolor=blue}\href{https://ui.adsabs.harvard.edu/abs/2021ApJ...923..265K}{923,
  265}

\bibitem[{Khokhlov(1991)}]{khokhlov_delayed_1991}
Khokhlov, A.~M. 1991, A\&A,
  \hypersetup{urlcolor=blue}\href{https://ui.adsabs.harvard.edu/abs/1991A&A...245..114K}{245,
  114}

\bibitem[{Landolt(1983)}]{landolt_ubvri_1983}
Landolt, A.~U. 1983,
  \hypersetup{urlcolor=magenta}\href{https://dx.doi.org/10.1086/113329}{AJ},
  \hypersetup{urlcolor=blue}\href{https://ui.adsabs.harvard.edu/abs/1983AJ.....88..439L}{88,
  439}

\bibitem[{Landolt(1992)}]{landolt_ubvri_1992}
Landolt, A.~U. 1992,
  \hypersetup{urlcolor=magenta}\href{https://dx.doi.org/10.1086/116242}{AJ},
  \hypersetup{urlcolor=blue}\href{https://ui.adsabs.harvard.edu/abs/1992AJ....104..340L}{104,
  340}

\bibitem[{Lang {et~al.}(2010)Lang, Hogg, Mierle, Blanton, \&
  Roweis}]{lang_astrometry_2010}
Lang, D., Hogg, D.~W., Mierle, K., Blanton, M., \& Roweis, S. 2010,
  \hypersetup{urlcolor=magenta}\href{https://dx.doi.org/10.1088/0004-6256/139/5/1782}{AJ},
  \hypersetup{urlcolor=blue}\href{https://ui.adsabs.harvard.edu/abs/2010AJ....139.1782L}{139,
  1782}

\bibitem[{Leonard(2007)}]{leonard_constraining_2007}
Leonard, D.~C. 2007,
  \hypersetup{urlcolor=magenta}\href{https://dx.doi.org/10.1086/522367}{ApJ},
  \hypersetup{urlcolor=blue}\href{https://ui.adsabs.harvard.edu/abs/2007ApJ...670.1275L}{670,
  1275}

\bibitem[{Leung {et~al.}(2023)Leung, Izzo, Hajela, Wang, Auchettl, Colle,
  Maeda, \& Murphy}]{leung_atca_2023}
Leung, J., Izzo, L., Hajela, A., {et~al.} 2023,
  \hypersetup{urlcolor=magenta}\href{http://astronomerstelegram.org/?read=15890}{ATel},
  \hypersetup{urlcolor=blue}\href{https://ui.adsabs.harvard.edu/abs/2023ATel.15890...1L}{15890,
  1}

\bibitem[{Li {et~al.}(2019)Li, Wang, Vink{\'o}, Mo, Hosseinzadeh, Sand, Zhang,
  Lin, Zhang, Wang, Zhang, Chen, Xiang, Rui, Huang, Li, Zhang, Li, Baron,
  Derkacy, Zhao, Sai, Zhang, Wang, Howell, McCully, Arcavi, Valenti, Hiramatsu,
  Burke, Rest, Garnavich, Tucker, Narayan, Shaya, Margheim, Zenteno, Villar,
  Dimitriadis, Foley, Pan, Coulter, Fox, Jha, Jones, Kasen, Kilpatrick, Piro,
  Riess, {Rojas-Bravo}, Shappee, Holoien, Stanek, Drout, Auchettl, Kochanek,
  Brown, Bose, Bersier, Brimacombe, Chen, Dong, Holmbo, Mu{\~n}oz, Mutel, Post,
  Prieto, Shields, Tallon, Thompson, Vallely, Villanueva, Smartt, Smith,
  Chambers, Flewelling, Huber, Magnier, Waters, Schultz, Bulger, Lowe, Willman,
  S{\'a}rneczky, P{\'a}l, Wheeler, B{\'o}di, Bogn{\'a}r, Cs{\'a}k, Cseh,
  Cs{\"o}rnyei, Hanyecz, Ign{\'a}cz, Kalup, {K{\"o}nyves-T{\'o}th}, Kriskovics,
  Ordasi, Rajmon, S{\'o}dor, Szab{\'o}, Szak{\'a}ts, Zsidi, Milne, Andrews,
  Smith, Bilinski, Brown, Nordin, Williams, Galbany, Palmerio, Hook, Inserra,
  Maguire, Cartier, Razza, Guti{\'e}rrez, Hermes, Reding, Kaiser, Tonry,
  Heinze, Denneau, Weiland, Stalder, Barentsen, Dotson, Barclay,
  {Gully-Santiago}, Hedges, Cody, Howell, Coughlin, Cleve, Cardoso, Larson,
  {McCalmont-Everton}, Peterson, Ross, Reedy, Osborne, McGinn, Kohnert,
  Migliorini, Wheaton, Spencer, Labonde, Castillo, Beerman, Steward, Hanley,
  Larsen, Gangopadhyay, Kloetzel, Weschler, Nystrom, Moffatt, Redick, Griest,
  Packard, Muszynski, Kampmeier, Bjella, Flynn, Elsaesser, {and}, {and}, {and},
  {and}, {and}, {and}, \& {and}}]{li_photometric_2019}
Li, W., Wang, X., Vink{\'o}, J., {et~al.} 2019,
  \hypersetup{urlcolor=magenta}\href{https://dx.doi.org/10.3847/1538-4357/aaec74}{ApJ},
  \hypersetup{urlcolor=blue}\href{https://ui.adsabs.harvard.edu/abs/2019ApJ...870...12L}{870,
  12}

\bibitem[{Lim {et~al.}(2023)Lim, Im, Paek, Yoon, Choi, Kim, Wheeler, Thomas,
  Vink{\'o}, Kim, Seo, Kang, Kim, Sung, Kim, Yoon, Kim, Kim, Bae, Ehgamberdiev,
  Burhonov, \& Mirzaqulov}]{lim_early_2023}
Lim, G., Im, M., Paek, G. S.~H., {et~al.} 2023,
  \hypersetup{urlcolor=magenta}\href{https://dx.doi.org/10.3847/1538-4357/acc10c}{ApJ},
  \hypersetup{urlcolor=blue}\href{https://ui.adsabs.harvard.edu/abs/2023ApJ...949...33L}{949,
  33}

\bibitem[{Livne(1990)}]{livne_successive_1990}
Livne, E. 1990,
  \hypersetup{urlcolor=magenta}\href{https://dx.doi.org/10.1086/185721}{ApJL},
  \hypersetup{urlcolor=blue}\href{https://ui.adsabs.harvard.edu/abs/1990ApJL..354L..53L}{354,
  L53}

\bibitem[{Lundqvist {et~al.}(2013)Lundqvist, Mattila, Sollerman, Kozma, Baron,
  Cox, Fransson, Leibundgut, \& Spyromilio}]{lundqvist_hydrogen_2013}
Lundqvist, P., Mattila, S., Sollerman, J., {et~al.} 2013,
  \hypersetup{urlcolor=magenta}\href{https://dx.doi.org/10.1093/mnras/stt1303}{MNRAS},
  \hypersetup{urlcolor=blue}\href{https://ui.adsabs.harvard.edu/abs/2013MNRAS.435..329L}{435,
  329}

\bibitem[{Lundqvist {et~al.}(2020)Lundqvist, Kundu, {P{\'e}rez-Torres}, Ryder,
  Bj{\"o}rnsson, Moldon, Argo, Beswick, Alberdi, \&
  Kool}]{lundqvist_deepest_2020}
Lundqvist, P., Kundu, E., {P{\'e}rez-Torres}, M.~A., {et~al.} 2020,
  \hypersetup{urlcolor=magenta}\href{https://dx.doi.org/10.3847/1538-4357/ab6dc6}{ApJ},
  \hypersetup{urlcolor=blue}\href{https://ui.adsabs.harvard.edu/abs/2020ApJ...890..159L}{890,
  159}

\bibitem[{Magee \& Maguire(2020)}]{magee_investigation_2020}
Magee, M.~R., \& Maguire, K. 2020,
  \hypersetup{urlcolor=magenta}\href{https://dx.doi.org/10.1051/0004-6361/202037870}{A\&A},
  \hypersetup{urlcolor=blue}\href{https://ui.adsabs.harvard.edu/abs/2020A&A...642A.189M}{642,
  A189}

\bibitem[{Maguire {et~al.}(2014)Maguire, Sullivan, Pan, {Gal-Yam}, Hook,
  Howell, Nugent, Mazzali, Chotard, Clubb, Filippenko, Kasliwal, Kandrashoff,
  Poznanski, Saunders, Silverman, Walker, \& Xu}]{maguire_exploring_2014}
Maguire, K., Sullivan, M., Pan, Y.-C., {et~al.} 2014,
  \hypersetup{urlcolor=magenta}\href{https://dx.doi.org/10.1093/mnras/stu1607}{MNRAS},
  \hypersetup{urlcolor=blue}\href{https://ui.adsabs.harvard.edu/abs/2014MNRAS.444.3258M}{444,
  3258}

\bibitem[{Marion {et~al.}(2016)Marion, Brown, Vink{\'o}, Silverman, Sand,
  Challis, Kirshner, Wheeler, Berlind, Brown, Calkins, Camacho, Dhungana,
  Foley, Friedman, Graham, Howell, Hsiao, Irwin, Jha, Kehoe, Macri, Maeda,
  Mandel, McCully, Pandya, Rines, Wilhelmy, \& Zheng}]{marion_sn_2016}
Marion, G.~H., Brown, P.~J., Vink{\'o}, J., {et~al.} 2016,
  \hypersetup{urlcolor=magenta}\href{https://dx.doi.org/10.3847/0004-637X/820/2/92}{ApJ},
  \hypersetup{urlcolor=blue}\href{https://ui.adsabs.harvard.edu/abs/2016ApJ...820...92M}{820,
  92}

\bibitem[{McCully {et~al.}(2018)McCully, Turner, Volgenau, Harbeck, Valenti,
  Riba, Bachelet, Snyder, Kurczynski, Norbury, \& Street}]{mccully_lcogt_2018}
McCully, C., Turner, M., Volgenau, N., {et~al.} 2018, {{LCOGT}}/Banzai:
  {{Initial Release}} v0.9.4, Zenodo,
  \hypersetup{urlcolor=magenta}doi:\href{https://doi.org/10.5281/zenodo.1257560}{10.5281/zenodo.1257560}

\bibitem[{McMullin {et~al.}(2007)McMullin, Waters, Schiebel, Young, \&
  Golap}]{mcmullin_casa_2007}
McMullin, J.~P., Waters, B., Schiebel, D., Young, W., \& Golap, K. 2007, ASPC,
  \hypersetup{urlcolor=blue}\href{https://ui.adsabs.harvard.edu/abs/2007ASPC..376..127M}{376,
  127}

\bibitem[{Miller {et~al.}(2018)Miller, Cao, Piro, Blagorodnova, Bue, Cenko,
  Dhawan, Ferretti, Fox, Fremling, Goobar, Howell, Hosseinzadeh, Kasliwal,
  Laher, Lunnan, Masci, McCully, Nugent, Sollerman, Taddia, \&
  Kulkarni}]{miller_early_2018}
Miller, A.~A., Cao, Y., Piro, A.~L., {et~al.} 2018,
  \hypersetup{urlcolor=magenta}\href{https://dx.doi.org/10.3847/1538-4357/aaa01f}{ApJ},
  \hypersetup{urlcolor=blue}\href{https://ui.adsabs.harvard.edu/abs/2018ApJ...852..100M}{852,
  100}

\bibitem[{Miller {et~al.}(2020)Miller, Magee, Polin, Maguire, Zimmerman, Yao,
  Sollerman, Schulze, Perley, Kromer, Dhawan, Bulla, Andreoni, Bellm, De,
  Dekany, Delacroix, Fremling, {Gal-Yam}, Goldstein, Golkhou, Goobar, Graham,
  Irani, Kasliwal, Kaye, Kim, Laher, Mahabal, Masci, Nugent, Ofek, Phinney,
  Prentice, Riddle, Rigault, Rusholme, Schweyer, Shupe, Soumagnac, Terreran,
  Walters, Yan, Zolkower, \& Kulkarni}]{miller_spectacular_2020}
Miller, A.~A., Magee, M.~R., Polin, A., {et~al.} 2020,
  \hypersetup{urlcolor=magenta}\href{https://dx.doi.org/10.3847/1538-4357/ab9e05}{ApJ},
  \hypersetup{urlcolor=blue}\href{https://ui.adsabs.harvard.edu/abs/2020ApJ...898...56M}{898,
  56}

\bibitem[{{NASA HEASARC}(2014)}]{nasaheasarc_heasoft:_2014}
{NASA HEASARC}. 2014, {{HEAsoft}}: {{Unified Release}} of {{FTOOLS}} and
  {{XANADU}},
  \hypersetup{urlcolor=magenta}\href{https://ascl.net/1408.004}{ascl}:\hypersetup{urlcolor=blue}\href{https://ui.adsabs.harvard.edu/abs/2014ascl.soft08004N}{1408.004}

\bibitem[{{National Optical Astronomy
  Observatories}(1999)}]{nationalopticalastronomyobservatories_iraf:_1999}
{National Optical Astronomy Observatories}. 1999, {{IRAF}}: {{Image}} Reduction
  and Analysis Facility,
  \hypersetup{urlcolor=magenta}\href{https://ascl.net/9911.002}{ascl}:\hypersetup{urlcolor=blue}\href{https://ui.adsabs.harvard.edu/abs/1999ascl.soft11002N}{9911.002}

\bibitem[{Ni {et~al.}(2023{\natexlab{\hspace{0pt}a}})Ni, Moon, Drout, Matzner,
  Leong, Kim, Park, \& Lee}]{ni_origin_2023}
Ni, Y.~Q., Moon, D.-S., Drout, M.~R., {et~al.} 2023{\natexlab{\hspace{0pt}a}},
  \hypersetup{urlcolor=magenta}\href{https://arxiv.org/abs/2304.00625}{arXiv}{:}\hypersetup{urlcolor=blue}\href{https://ui.adsabs.harvard.edu/abs/2023arXiv230400625N}{2304.00625}

\bibitem[{Ni {et~al.}(2022)Ni, Moon, Drout, Polin, Sand,
  {Gonz{\'a}lez-Gait{\'a}n}, Kim, Lee, Park, Howell, Nugent, Piro, Brown,
  Galbany, Burke, Hiramatsu, Hosseinzadeh, Valenti, Afsariardchi, Andrews,
  Antoniadis, Arcavi, Beaton, Bostroem, Carlberg, Cenko, Cha, Dong, {Gal-Yam},
  Haislip, Holoien, Johnson, Kouprianov, Lee, Matzner, Morrell, McCully,
  Pignata, Reichart, Rich, Ryder, Smith, Wyatt, \& Yang}]{ni_infant-phase_2022}
Ni, Y.~Q., Moon, D.-S., Drout, M.~R., {et~al.} 2022,
  \hypersetup{urlcolor=magenta}\href{https://dx.doi.org/10.1038/s41550-022-01603-4}{NatAs},
  \hypersetup{urlcolor=blue}\href{https://ui.adsabs.harvard.edu/abs/2022NatAs...6..568N}{6,
  568}

\bibitem[{Ni {et~al.}(2023{\natexlab{\hspace{0pt}b}})Ni, Moon, Drout, Polin,
  Sand, {Gonz{\'a}lez-Gait{\'a}n}, Kim, Lee, Park, Howell, Nugent, Piro, Brown,
  Galbany, Burke, Hiramatsu, Hosseinzadeh, Valenti, Afsariardchi, Andrews,
  Antoniadis, Beaton, Bostroem, Carlberg, Cenko, Cha, Dong, {Gal-Yam}, Haislip,
  Holoien, Johnson, Kouprianov, Lee, Matzner, Morrell, McCully, Pignata,
  Reichart, Rich, Ryder, Smith, Wyatt, \& Yang}]{ni_origin_2023a}
Ni, Y.~Q., Moon, D.-S., Drout, M.~R., {et~al.} 2023{\natexlab{\hspace{0pt}b}},
  \hypersetup{urlcolor=magenta}\href{https://dx.doi.org/10.3847/1538-4357/aca9be}{ApJ},
  \hypersetup{urlcolor=blue}\href{https://ui.adsabs.harvard.edu/abs/2023ApJ...946....7N}{946,
  7}

\bibitem[{Noebauer {et~al.}(2017)Noebauer, Kromer, Taubenberger, Baklanov,
  Blinnikov, Sorokina, \& Hillebrandt}]{noebauer_early_2017}
Noebauer, U.~M., Kromer, M., Taubenberger, S., {et~al.} 2017,
  \hypersetup{urlcolor=magenta}\href{https://dx.doi.org/10.1093/mnras/stx2093}{MNRAS},
  \hypersetup{urlcolor=blue}\href{https://ui.adsabs.harvard.edu/abs/2017MNRAS.472.2787N}{472,
  2787}

\bibitem[{Nomoto(1982)}]{nomoto_accreting_1982}
Nomoto, K. 1982,
  \hypersetup{urlcolor=magenta}\href{https://dx.doi.org/10.1086/160031}{ApJ},
  \hypersetup{urlcolor=blue}\href{https://ui.adsabs.harvard.edu/abs/1982ApJ...257..780N}{257,
  780}

\bibitem[{Nugent {et~al.}(2011)Nugent, Sullivan, Cenko, Thomas, Kasen, Howell,
  Bersier, Bloom, Kulkarni, Kandrashoff, Filippenko, Silverman, Marcy, Howard,
  Isaacson, Maguire, Suzuki, Tarlton, Pan, Bildsten, Fulton, Parrent, Sand,
  Podsiadlowski, Bianco, Dilday, Graham, Lyman, James, Kasliwal, Law, Quimby,
  Hook, Walker, Mazzali, Pian, Ofek, {Gal-Yam}, \&
  Poznanski}]{nugent_supernova_2011}
Nugent, P.~E., Sullivan, M., Cenko, S.~B., {et~al.} 2011,
  \hypersetup{urlcolor=magenta}\href{https://dx.doi.org/10.1038/nature10644}{Natur},
  \hypersetup{urlcolor=blue}\href{https://ui.adsabs.harvard.edu/abs/2011Natur.480..344N}{480,
  344}

\bibitem[{Oliphant(2006)}]{oliphant_guide_2006}
Oliphant, T.~E. 2006, A Guide to {{NumPy}} ({USA}: {Trelgol Publishing})

\bibitem[{Olling {et~al.}(2015)Olling, Mushotzky, Shaya, Rest, Garnavich,
  Tucker, Kasen, Margheim, \& Filippenko}]{olling_no_2015}
Olling, R.~P., Mushotzky, R., Shaya, E.~J., {et~al.} 2015,
  \hypersetup{urlcolor=magenta}\href{https://dx.doi.org/10.1038/nature14455}{Natur},
  \hypersetup{urlcolor=blue}\href{https://ui.adsabs.harvard.edu/abs/2015Natur.521..332O}{521,
  332}

\bibitem[{Parrent {et~al.}(2014)Parrent, Friesen, \&
  Parthasarathy}]{parrent_review_2014}
Parrent, J., Friesen, B., \& Parthasarathy, M. 2014,
  \hypersetup{urlcolor=magenta}\href{https://dx.doi.org/10.1007/s10509-014-1830-1}{Ap\&SS},
  \hypersetup{urlcolor=blue}\href{https://ui.adsabs.harvard.edu/abs/2014Ap&SS.351....1P}{351,
  1}

\bibitem[{Parrent {et~al.}(2011)Parrent, Thomas, Fesen, Marion, Challis,
  Garnavich, {Dan Milisavljevic}, Vink{\`o}, \& Wheeler}]{parrent_study_2011}
Parrent, J.~T., Thomas, R.~C., Fesen, R.~A., {et~al.} 2011,
  \hypersetup{urlcolor=magenta}\href{https://dx.doi.org/10.1088/0004-637X/732/1/30}{ApJ},
  \hypersetup{urlcolor=blue}\href{https://ui.adsabs.harvard.edu/abs/2011ApJ...732...30P}{732,
  30}

\bibitem[{Pellegrino {et~al.}(2020)Pellegrino, Howell, Sarbadhicary, Burke,
  Hiramatsu, McCully, Milne, Andrews, Brown, Chomiuk, Hsiao, Sand, Shahbandeh,
  Smith, Valenti, Vink{\'o}, Wheeler, Wyatt, \&
  Yang}]{pellegrino_constraining_2020}
Pellegrino, C., Howell, D.~A., Sarbadhicary, S.~K., {et~al.} 2020,
  \hypersetup{urlcolor=magenta}\href{https://dx.doi.org/10.3847/1538-4357/ab8e3f}{ApJ},
  \hypersetup{urlcolor=blue}\href{https://ui.adsabs.harvard.edu/abs/2020ApJ...897..159P}{897,
  159}

\bibitem[{{P{\'e}rez-Torres} {et~al.}(2014){P{\'e}rez-Torres}, Lundqvist,
  Beswick, Bj{\"o}rnsson, Muxlow, Paragi, Ryder, Alberdi, Fransson, Marcaide,
  {Mart{\'i}-Vidal}, Ros, Argo, \& Guirado}]{perez-torres_constraints_2014}
{P{\'e}rez-Torres}, M.~A., Lundqvist, P., Beswick, R.~J., {et~al.} 2014,
  \hypersetup{urlcolor=magenta}\href{https://dx.doi.org/10.1088/0004-637X/792/1/38}{ApJ},
  \hypersetup{urlcolor=blue}\href{https://ui.adsabs.harvard.edu/abs/2014ApJ...792...38P}{792,
  38}

\bibitem[{Phillips(1993)}]{phillips_absolute_1993}
Phillips, M.~M. 1993,
  \hypersetup{urlcolor=magenta}\href{https://dx.doi.org/10.1086/186970}{ApJL},
  \hypersetup{urlcolor=blue}\href{https://ui.adsabs.harvard.edu/abs/1993ApJL..413L.105P}{413,
  L105}

\bibitem[{Phillips {et~al.}(2013)Phillips, Simon, Morrell, Burns, Cox, Foley,
  Karakas, Patat, Sternberg, Williams, {Gal-Yam}, Hsiao, Leonard, Persson,
  Stritzinger, Thompson, Campillay, Contreras, Folatelli, Freedman, Hamuy,
  Roth, Shields, Suntzeff, Chomiuk, Ivans, Madore, Penprase, Perley, Pignata,
  Preston, \& Soderberg}]{phillips_source_2013}
Phillips, M.~M., Simon, J.~D., Morrell, N., {et~al.} 2013,
  \hypersetup{urlcolor=magenta}\href{https://dx.doi.org/10.1088/0004-637X/779/1/38}{ApJ},
  \hypersetup{urlcolor=blue}\href{https://ui.adsabs.harvard.edu/abs/2013ApJ...779...38P}{779,
  38}

\bibitem[{Piro \& Morozova(2016)}]{piro_exploring_2016}
Piro, A.~L., \& Morozova, V.~S. 2016,
  \hypersetup{urlcolor=magenta}\href{https://dx.doi.org/10.3847/0004-637X/826/1/96}{ApJ},
  \hypersetup{urlcolor=blue}\href{https://ui.adsabs.harvard.edu/abs/2016ApJ...826...96P}{826,
  96}

\bibitem[{Polin {et~al.}(2019)Polin, Nugent, \&
  Kasen}]{polin_observational_2019}
Polin, A., Nugent, P., \& Kasen, D. 2019,
  \hypersetup{urlcolor=magenta}\href{https://dx.doi.org/10.3847/1538-4357/aafb6a}{ApJ},
  \hypersetup{urlcolor=blue}\href{https://ui.adsabs.harvard.edu/abs/2019ApJ...873...84P}{873,
  84}

\bibitem[{Poznanski {et~al.}(2011)Poznanski, Ganeshalingam, Silverman, \&
  Filippenko}]{poznanski_lowresolution_2011}
Poznanski, D., Ganeshalingam, M., Silverman, J.~M., \& Filippenko, A.~V. 2011,
  \hypersetup{urlcolor=magenta}\href{https://dx.doi.org/10.1111/j.1745-3933.2011.01084.x}{MNRAS},
  \hypersetup{urlcolor=blue}\href{https://ui.adsabs.harvard.edu/abs/2011MNRAS.415L..81P}{415,
  L81}

\bibitem[{Poznanski {et~al.}(2012)Poznanski, Prochaska, \&
  Bloom}]{poznanski_empirical_2012}
Poznanski, D., Prochaska, J.~X., \& Bloom, J.~S. 2012,
  \hypersetup{urlcolor=magenta}\href{https://dx.doi.org/10.1111/j.1365-2966.2012.21796.x}{MNRAS},
  \hypersetup{urlcolor=blue}\href{https://ui.adsabs.harvard.edu/abs/2012MNRAS.426.1465P}{426,
  1465}

\bibitem[{Rau \& Cornwell(2011)}]{rau_multiscale_2011}
Rau, U., \& Cornwell, T.~J. 2011,
  \hypersetup{urlcolor=magenta}\href{https://dx.doi.org/10.1051/0004-6361/201117104}{A\&A},
  \hypersetup{urlcolor=blue}\href{https://ui.adsabs.harvard.edu/abs/2011A&A...532A..71R}{532,
  A71}

\bibitem[{Reichart {et~al.}(2005)Reichart, Nysewander, Moran, Bartelme,
  Bayliss, Foster, Clemens, Price, Evans, Salmonson, Trammell, Carney, Keohane,
  \& Gotwals}]{reichart_prompt:_2005}
Reichart, D., Nysewander, M., Moran, J., {et~al.} 2005,
  \hypersetup{urlcolor=magenta}\href{https://dx.doi.org/10.1393/ncc/i2005-10149-6}{NCimC},
  \hypersetup{urlcolor=blue}\href{https://ui.adsabs.harvard.edu/abs/2005NCimC..28..767R}{28,
  767}

\bibitem[{Robitaille {et~al.}(2023)Robitaille, Deil, \&
  Ginsburg}]{robitaille_image_2023}
Robitaille, T., Deil, C., \& Ginsburg, A. 2023, Image Reprojection
  (Resampling),
  \hypersetup{urlcolor=magenta}\url{https://reproject.readthedocs.io/}

\bibitem[{Roming {et~al.}(2005)Roming, Kennedy, Mason, Nousek, Ahr, Bingham,
  Broos, Carter, Hancock, Huckle, Hunsberger, Kawakami, Killough, Koch,
  Mclelland, Smith, Smith, Soto, Boyd, Breeveld, Holland, Ivanushkina, Pryzby,
  Still, \& Stock}]{roming_swift_2005}
Roming, P. W.~A., Kennedy, T.~E., Mason, K.~O., {et~al.} 2005,
  \hypersetup{urlcolor=magenta}\href{https://dx.doi.org/10.1007/s11214-005-5095-4}{SSRv},
  \hypersetup{urlcolor=blue}\href{https://ui.adsabs.harvard.edu/abs/2005SSRv..120...95R}{120,
  95}

\bibitem[{Rubin {et~al.}(1974)Rubin, Westpfahl, \& Tuve}]{rubin_second_1974}
Rubin, V.~C., Westpfahl, D., \& Tuve, M. 1974,
  \hypersetup{urlcolor=magenta}\href{https://dx.doi.org/10.1086/111692}{AJ},
  \hypersetup{urlcolor=blue}\href{https://ui.adsabs.harvard.edu/abs/1974AJ.....79.1406R}{79,
  1406}

\bibitem[{Sand {et~al.}(2018)Sand, Graham, Boty{\'a}nszki, Hiramatsu, McCully,
  Valenti, Hosseinzadeh, Howell, Burke, Cartier, Diamond, Hsiao, Jha, Kasen,
  Kumar, Marion, Suntzeff, Tartaglia, Wheeler, \& Wyatt}]{sand_nebular_2018}
Sand, D.~J., Graham, M.~L., Boty{\'a}nszki, J., {et~al.} 2018,
  \hypersetup{urlcolor=magenta}\href{https://dx.doi.org/10.3847/1538-4357/aacde8}{ApJ},
  \hypersetup{urlcolor=blue}\href{https://ui.adsabs.harvard.edu/abs/2018ApJ...863...24S}{863,
  24}

\bibitem[{Sand {et~al.}(2019)Sand, Amaro, Moe, Graham, Andrews, Burke, Cartier,
  Eweis, Galbany, Hiramatsu, Howell, Jha, Lundquist, Matheson, McCully, Milne,
  Smith, Valenti, \& Wyatt}]{sand_nebular_2019}
Sand, D.~J., Amaro, R.~C., Moe, M., {et~al.} 2019,
  \hypersetup{urlcolor=magenta}\href{https://dx.doi.org/10.3847/2041-8213/ab1eaf}{ApJL},
  \hypersetup{urlcolor=blue}\href{https://ui.adsabs.harvard.edu/abs/2019ApJL..877L...4S}{877,
  L4}

\bibitem[{Schlafly \& Finkbeiner(2011)}]{schlafly_measuring_2011}
Schlafly, E.~F., \& Finkbeiner, D.~P. 2011,
  \hypersetup{urlcolor=magenta}\href{https://dx.doi.org/10.1088/0004-637X/737/2/103}{ApJ},
  \hypersetup{urlcolor=blue}\href{https://ui.adsabs.harvard.edu/abs/2011ApJ...737..103S}{737,
  103}

\bibitem[{{Science Software Branch at
  STScI}(2012)}]{sciencesoftwarebranchatstsci_pyraf:_2012}
{Science Software Branch at STScI}. 2012, {{PyRAF}}: {{Python}} Alternative for
  {{IRAF}},
  \hypersetup{urlcolor=magenta}\href{https://ascl.net/1207.011}{ascl}:\hypersetup{urlcolor=blue}\href{https://ui.adsabs.harvard.edu/abs/2012ascl.soft07011S}{1207.011}

\bibitem[{Seaquist \& Taylor(1990)}]{seaquist_collective_1990}
Seaquist, E.~R., \& Taylor, A.~R. 1990,
  \hypersetup{urlcolor=magenta}\href{https://dx.doi.org/10.1086/168315}{ApJ},
  \hypersetup{urlcolor=blue}\href{https://ui.adsabs.harvard.edu/abs/1990ApJ...349..313S}{349,
  313}

\bibitem[{Shappee {et~al.}(2019)Shappee, Holoien, Drout, Auchettl, Stritzinger,
  Kochanek, Stanek, Shaya, Narayan, Brown, Bose, Bersier, Brimacombe, Chen,
  Dong, Holmbo, Katz, Mu{\~n}oz, Mutel, Post, Prieto, Shields, Tallon,
  Thompson, Vallely, Villanueva, Denneau, Flewelling, Heinze, Smith, Stalder,
  Tonry, Weiland, Barclay, Barentsen, Cody, Dotson, Foerster, Garnavich,
  {Gully-Santiago}, Hedges, Howell, Kasen, Margheim, Mushotzky, Rest, Tucker,
  Villar, Zenteno, Beerman, Bjella, Castillo, Coughlin, Elsaesser, Flynn,
  Gangopadhyay, Griest, Hanley, Kampmeier, Kloetzel, Kohnert, Labonde, Larsen,
  Larson, {McCalmont-Everton}, McGinn, Migliorini, Moffatt, Muszynski, Nystrom,
  Osborne, Packard, Peterson, Redick, Reedy, Ross, Spencer, Steward, Cleve,
  Cardoso, Weschler, Wheaton, Bulger, Chambers, Flewelling, Huber, Lowe,
  Magnier, Schultz, Waters, Willman, Baron, Chen, Derkacy, Huang, Li, Li, Li,
  Mo, Rui, Sai, Wang, Wang, Wang, Xiang, Zhang, Zhang, Zhang, Zhang, Zhang,
  Zhao, Brown, Hermes, Nordin, Points, S{\'o}dor, Strampelli, Zenteno, {and},
  {and}, \& {and}}]{shappee_seeing_2019}
Shappee, B.~J., Holoien, T. W.-S., Drout, M.~R., {et~al.} 2019,
  \hypersetup{urlcolor=magenta}\href{https://dx.doi.org/10.3847/1538-4357/aaec79}{ApJ},
  \hypersetup{urlcolor=blue}\href{https://ui.adsabs.harvard.edu/abs/2019ApJ...870...13S}{870,
  13}

\bibitem[{Shen {et~al.}(2021)Shen, Boos, Townsley, \&
  Kasen}]{shen_multidimensional_2021}
Shen, K.~J., Boos, S.~J., Townsley, D.~M., \& Kasen, D. 2021,
  \hypersetup{urlcolor=magenta}\href{https://dx.doi.org/10.3847/1538-4357/ac2304}{ApJ},
  \hypersetup{urlcolor=blue}\href{https://ui.adsabs.harvard.edu/abs/2021ApJ...922...68S}{922,
  68}

\bibitem[{Shen {et~al.}(2018)Shen, Boubert, G{\"a}nsicke, Jha, Andrews,
  Chomiuk, Foley, Fraser, Gromadzki, Guillochon, Kotze, Maguire, Siebert,
  Smith, Strader, Badenes, Kerzendorf, Koester, Kromer, Miles, Pakmor, Schwab,
  Toloza, Toonen, Townsley, \& Williams}]{shen_three_2018}
Shen, K.~J., Boubert, D., G{\"a}nsicke, B.~T., {et~al.} 2018,
  \hypersetup{urlcolor=magenta}\href{https://dx.doi.org/10.3847/1538-4357/aad55b}{ApJ},
  \hypersetup{urlcolor=blue}\href{https://ui.adsabs.harvard.edu/abs/2018ApJ...865...15S}{865,
  15}

\bibitem[{Siebert {et~al.}(2020)Siebert, Dimitriadis, Polin, \&
  Foley}]{siebert_strong_2020}
Siebert, M.~R., Dimitriadis, G., Polin, A., \& Foley, R.~J. 2020,
  \hypersetup{urlcolor=magenta}\href{https://dx.doi.org/10.3847/2041-8213/abae6e}{ApJL},
  \hypersetup{urlcolor=blue}\href{https://ui.adsabs.harvard.edu/abs/2020ApJL..900L..27S}{900,
  L27}

\bibitem[{Silverman \& Filippenko(2012)}]{silverman_berkeley_2012a}
Silverman, J.~M., \& Filippenko, A.~V. 2012,
  \hypersetup{urlcolor=magenta}\href{https://dx.doi.org/10.1111/j.1365-2966.2012.21276.x}{MNRAS},
  \hypersetup{urlcolor=blue}\href{https://ui.adsabs.harvard.edu/abs/2012MNRAS.425.1917S}{425,
  1917}

\bibitem[{Silverman {et~al.}(2013)Silverman, Nugent, {Gal-Yam}, Sullivan,
  Howell, Filippenko, Arcavi, {Ben-Ami}, Bloom, Cenko, Cao, Chornock, Clubb,
  Coil, Foley, Graham, Griffith, Horesh, Kasliwal, Kulkarni, Leonard, Li,
  Matheson, Miller, Modjaz, Ofek, Pan, Perley, Poznanski, Quimby, Steele,
  Sternberg, Xu, \& Yaron}]{silverman_type_2013}
Silverman, J.~M., Nugent, P.~E., {Gal-Yam}, A., {et~al.} 2013,
  \hypersetup{urlcolor=magenta}\href{https://dx.doi.org/10.1088/0067-0049/207/1/3}{ApJS},
  \hypersetup{urlcolor=blue}\href{https://ui.adsabs.harvard.edu/abs/2013ApJS..207....3S}{207,
  3}

\bibitem[{Smith {et~al.}(2006)Smith, Nordsieck, Burgh, Percival, Williams,
  O'Donohue, O'Connor, \& Schier}]{smith_prime_2006}
Smith, M.~P., Nordsieck, K.~H., Burgh, E.~B., {et~al.} 2006,
  \hypersetup{urlcolor=magenta}\href{https://dx.doi.org/10.1117/12.672415}{Proc.\
  SPIE},
  \hypersetup{urlcolor=blue}\href{https://ui.adsabs.harvard.edu/abs/2006SPIE.6269E..2AS}{6269,
  2A}

\bibitem[{Springob {et~al.}(2009)Springob, Masters, Haynes, Giovanelli, \&
  Marinoni}]{springob_erratum_2009}
Springob, C.~M., Masters, K.~L., Haynes, M.~P., Giovanelli, R., \& Marinoni, C.
  2009,
  \hypersetup{urlcolor=magenta}\href{https://dx.doi.org/10.1088/0067-0049/182/1/474}{ApJS},
  \hypersetup{urlcolor=blue}\href{https://ui.adsabs.harvard.edu/abs/2009ApJS..182..474S}{182,
  474}

\bibitem[{Tartaglia {et~al.}(2018)Tartaglia, Sand, Valenti, Wyatt, Anderson,
  Arcavi, Ashall, Botticella, {R. Cartier}, Chen, Cikota, Coulter, Valle,
  Foley, {Gal-Yam}, Galbany, Gall, Haislip, Harmanen, Hosseinzadeh, Howell,
  Hsiao, Inserra, Jha, Kankare, Kilpatrick, Kouprianov, Kuncarayakti,
  Maccarone, Maguire, Mattila, Mazzali, McCully, Melandri, Morrell, Phillips,
  Pignata, Piro, Prentice, Reichart, {Rojas-Bravo}, Smartt, Smith, Sollerman,
  Stritzinger, Sullivan, {F. Taddia}, \& Young}]{tartaglia_early_2018}
Tartaglia, L., Sand, D.~J., Valenti, S., {et~al.} 2018,
  \hypersetup{urlcolor=magenta}\href{https://dx.doi.org/10.3847/1538-4357/aaa014}{ApJ},
  \hypersetup{urlcolor=blue}\href{https://ui.adsabs.harvard.edu/abs/2018ApJ...853...62T}{853,
  62}

\bibitem[{Thomas {et~al.}(2011)Thomas, Aldering, Antilogus, Aragon, Bailey,
  Baltay, Bongard, Buton, Canto, Childress, Chotard, Copin, Fakhouri, Gangler,
  Hsiao, Kerschhaggl, Kowalski, Loken, Nugent, Paech, Pain, Pecontal, Pereira,
  Perlmutter, Rabinowitz, Rigault, Rubin, Runge, Scalzo, Smadja, Tao, Weaver,
  Wu, Factory), Brown, \& Milne}]{thomas_type_2011}
Thomas, R.~C., Aldering, G., Antilogus, P., {et~al.} 2011,
  \hypersetup{urlcolor=magenta}\href{https://dx.doi.org/10.1088/0004-637X/743/1/27}{ApJ},
  \hypersetup{urlcolor=blue}\href{https://ui.adsabs.harvard.edu/abs/2011ApJ...743...27T}{743,
  27}

\bibitem[{Torres {et~al.}(2017)Torres, Brice{\~n}o, \&
  Quint}]{torres_goodman_2017}
Torres, S., Brice{\~n}o, C., \& Quint, B. 2017, Goodman {{HTS Pipeline
  Documentation}} 1.3.6,
  \hypersetup{urlcolor=magenta}\url{https://soardocs.readthedocs.io/projects/goodman-pipeline/}

\bibitem[{Tucker {et~al.}(2019)Tucker, Shappee, \& Wisniewski}]{tucker_no_2019}
Tucker, M.~A., Shappee, B.~J., \& Wisniewski, J.~P. 2019,
  \hypersetup{urlcolor=magenta}\href{https://dx.doi.org/10.3847/2041-8213/ab0286}{ApJL},
  \hypersetup{urlcolor=blue}\href{https://ui.adsabs.harvard.edu/abs/2019ApJ...872L..22T}{872,
  L22}

\bibitem[{Tucker {et~al.}(2020)Tucker, Shappee, Vallely, Stanek, Prieto,
  Botyanszki, Kochanek, Anderson, Brown, Galbany, Holoien, Hsiao, Kumar,
  Kuncarayakti, Morrell, Phillips, Stritzinger, \&
  Thompson}]{tucker_nebular_2020}
Tucker, M.~A., Shappee, B.~J., Vallely, P.~J., {et~al.} 2020,
  \hypersetup{urlcolor=magenta}\href{https://dx.doi.org/10.1093/mnras/stz3390}{MNRAS},
  \hypersetup{urlcolor=blue}\href{https://ui.adsabs.harvard.edu/abs/2020MNRAS.493.1044T}{493,
  1044}

\bibitem[{Tucker {et~al.}(2021)Tucker, Ashall, Shappee, Vallely, Kochanek,
  Huber, Anand, Keane, Hsiao, \& Holoien}]{tucker_sn_2021}
Tucker, M.~A., Ashall, C., Shappee, B.~J., {et~al.} 2021,
  \hypersetup{urlcolor=magenta}\href{https://dx.doi.org/10.3847/1538-4357/abf93b}{ApJ},
  \hypersetup{urlcolor=blue}\href{https://ui.adsabs.harvard.edu/abs/2021ApJ...914...50T}{914,
  50}

\bibitem[{Tully \& Fisher(1977)}]{tully_new_1977}
Tully, R.~B., \& Fisher, J.~R. 1977, A\&A,
  \hypersetup{urlcolor=blue}\href{https://ui.adsabs.harvard.edu/abs/1977A&A....54..661T}{54,
  661}

\bibitem[{Valenti {et~al.}(2014)Valenti, Sand, Pastorello, Graham, Howell,
  Parrent, Tomasella, Ochner, Fraser, Benetti, Yuan, Smartt, Maund, Arcavi,
  {Gal-Yam}, Inserra, \& Young}]{valenti_first_2014}
Valenti, S., Sand, D., Pastorello, A., {et~al.} 2014,
  \hypersetup{urlcolor=magenta}\href{https://dx.doi.org/10.1093/mnrasl/slt171}{MNRAS},
  \hypersetup{urlcolor=blue}\href{https://ui.adsabs.harvard.edu/abs/2014MNRAS.438L.101V}{438,
  L101}

\bibitem[{Valenti {et~al.}(2016)Valenti, Howell, Stritzinger, Graham,
  Hosseinzadeh, Arcavi, Bildsten, Jerkstrand, McCully, Pastorello, Piro, Sand,
  Smartt, Terreran, Baltay, Benetti, Brown, Filippenko, Fraser, Rabinowitz,
  Sullivan, \& Yuan}]{valenti_diversity_2016}
Valenti, S., Howell, D.~A., Stritzinger, M.~D., {et~al.} 2016,
  \hypersetup{urlcolor=magenta}\href{https://dx.doi.org/10.1093/mnras/stw870}{MNRAS},
  \hypersetup{urlcolor=blue}\href{https://ui.adsabs.harvard.edu/abs/2016MNRAS.459.3939V}{459,
  3939}

\bibitem[{Wang {et~al.}(2023)Wang, Rest, Dimitriadis, {Ridden-harper}, Siebert,
  Magee, Angus, Auchettl, Davis, Foley, Fox, Gomez, Jencson, Jones, Kilpatrick,
  Pierel, Piro, Polin, Politsch, {Rojas-bravo}, Shahbandeh, Villar, Zenati,
  Ashall, Chambers, Coulter, De~Boer, Dilullo, Gall, Gao, Hsiao, Huber, Izzo,
  Khetan, Lebaron, Magnier, Mandel, Mcgill, Miao, Pan, Stevens, Swift, Taggart,
  \& Yang}]{wang_flight_2023}
Wang, Q., Rest, A., Dimitriadis, G., {et~al.} 2023,
  \hypersetup{urlcolor=magenta}\href{https://arxiv.org/abs/2305.03779}{arXiv}{:}\hypersetup{urlcolor=blue}\href{https://ui.adsabs.harvard.edu/abs/2023arXiv230503779W}{2305.03779}

\bibitem[{Webbink(1984)}]{webbink_double_1984}
Webbink, R.~F. 1984,
  \hypersetup{urlcolor=magenta}\href{https://dx.doi.org/10.1086/161701}{ApJ},
  \hypersetup{urlcolor=blue}\href{https://ui.adsabs.harvard.edu/abs/1984ApJ...277..355W}{277,
  355}

\bibitem[{Whelan \& Iben(1973)}]{whelan_binaries_1973}
Whelan, J., \& Iben, I. 1973,
  \hypersetup{urlcolor=magenta}\href{https://dx.doi.org/10.1086/152565}{ApJ},
  \hypersetup{urlcolor=blue}\href{https://ui.adsabs.harvard.edu/abs/1973ApJ...186.1007W}{186,
  1007}

\bibitem[{Woosley \& Kasen(2011)}]{woosley_sub-chandrasekhar_2011}
Woosley, S.~E., \& Kasen, D. 2011,
  \hypersetup{urlcolor=magenta}\href{https://dx.doi.org/10.1088/0004-637X/734/1/38}{ApJ},
  \hypersetup{urlcolor=blue}\href{https://ui.adsabs.harvard.edu/abs/2011ApJ...734...38W}{734,
  38}

\bibitem[{Yaron \& {Gal-Yam}(2012)}]{yaron_wiserep_2012}
Yaron, O., \& {Gal-Yam}, A. 2012,
  \hypersetup{urlcolor=magenta}\href{https://dx.doi.org/10.1086/666656}{PASP},
  \hypersetup{urlcolor=blue}\href{https://ui.adsabs.harvard.edu/abs/2012PASP..124..668Y}{124,
  668}

\bibitem[{Zhai {et~al.}(2023)Zhai, Zhang, Li, \& Wang}]{zhai_lions_2023}
Zhai, Q., Zhang, J., Li, L., \& Wang, X. 2023,
  \hypersetup{urlcolor=magenta}\href{https://ui.adsabs.harvard.edu/abs/2023TNSCR.260....1Z}{TNSCR},
  \hypersetup{urlcolor=blue}\href{https://ui.adsabs.harvard.edu/abs/2023TNSCR.260....1Z}{260,
  1}

\bibitem[{Zhang {et~al.}(2016)Zhang, Wang, Zhang, Zhang, Ganeshalingam, Li,
  Filippenko, Zhao, Zheng, Bai, Chen, Chen, Huang, Mo, Rui, Song, Sai, Li,
  Wang, \& Wu}]{zhang_optical_2016}
Zhang, K., Wang, X., Zhang, J., {et~al.} 2016,
  \hypersetup{urlcolor=magenta}\href{https://dx.doi.org/10.3847/0004-637X/820/1/67}{ApJ},
  \hypersetup{urlcolor=blue}\href{https://ui.adsabs.harvard.edu/abs/2016ApJ...820...67Z}{820,
  67}

\end{thebibliography}
